\documentclass[useAMS,usenatbib]{mn2e}
\usepackage{times}
\usepackage{epsfig}
\usepackage{graphicx}
\usepackage[usenames,dvips]{color}
\def\apj{ ApJ}
\def\aap{ A\&A}
\def\mnras{MNRAS}

\def\araa{ARAA}
\def\aj{AJ}

\def\apjl{ ApJL}

\title[White dwarf binaries in globular clusters]{Formation and evolution of compact binaries in globular clusters: 
I.~Binaries with white dwarfs.}
\author[N Ivanova et al.]
{N.\ Ivanova $^1$\thanks{E-mail:nata@cita.utoronto.ca}, C.~O.~Heinke$^2$\thanks{Lindheimer Fellow}, F.~A.\ Rasio$^2$, R.~E.\ Taam$^2$,  
K.\ Belczynski$^{3}$\thanks{Tombaugh Fellow},
\& J.\ Fregeau$^{2}$ \\
$^1$Canadian Institute for Theoretical Astrophysics, University of Toronto, 60 St. George, Toronto, ON M5S 3H8, Canada\\
$^2$Northwestern University, Dept of Physics \& Astronomy,  2145 Sheridan Rd, Evanston, IL 60208, USA\\
$^3$New Mexico State University, Department of Astronomy,  1320 Frenger Mall, Las Cruces, New Mexico 88003-8001, USA
}

\begin{document}

\maketitle

\label{firstpage}

\begin{abstract}{
In this paper, the first of a series, we study the stellar dynamical  and evolutionary
processes leading to the formation of
compact binaries containing white dwarfs in dense globular clusters. 
We examine the processes leading to the creation of X-ray binaries such as  cataclysmic variables
and AM CVn systems. Using numerical simulations, we identify the  dominant formation channels
and we predict the expected numbers and characteristics of detectable  systems, emphasizing
how the cluster sources differ from the field population. We explore  the dependence of
formation rates on cluster properties and we explain in particular  why the distribution of
cataclysmic variables has only a weak dependence on cluster density.  We also discuss the
frequency of dwarf nova outbursts in globular clusters and their  connection with moderately
strong white dwarf magnetic fields. We examine the rate of Type Ia  supernovae via both single
and double degenerate channels in clusters and we argue that those  rates may contribute
to the total SN Ia rate in elliptical galaxies. Considering  coalescing white dwarf binaries we
discuss possible constraints on the common envelope evolution of  their progenitors and we derive
theoretical expectations for gravitational wave detection by LISA. 
}
\end{abstract}


\section{Introduction}
From the earliest observations of X-ray binaries in globular clusters (GCs) 
it has been noted that they must be very efficient sites for the production 
of compact binary systems \citep{Clark75}.
The key to the overabundance of compact binaries in clusters, as compared 
to the field, is close stellar encounters. 
The processes that influence the binary population in dense stellar environments 
include the destruction of wide binaries (``ionization''), hardening of close binaries,
physical collisions, 
and exchange interactions, through which low-mass companions tend to be replaced
by more massive participants in the encounter. 
As a result of these processes, in the dense cores of globular clusters, 
binaries are strongly depleted and their period distribution is very different 
from that of a field population \citep{Ivanova05}.  This effect is 
stronger for binaries  
including a compact object, 
like cataclysmic variables (CVs).

The issue of the dynamical formation of CVs has been extensively discussed. 
Considering the CV formation via tidal captures, 
\cite{Bailyn90_tc} showed that dynamical formation of CVs is not expected 
because more massive donors lead to unstable mass transfer. 
On the other hand, \cite{DiStefano94} predicted the existence of many CVs 
 formed via tidal captures, as many as an order of magnitude more than would be predicted by
standard binary evolution, making CVs a probe of the dynamical processes in the cluster.
Detection of CVs in globular clusters proved difficult \citep[e.g.][]{Shara96}, 
but a population was detected using the {\it Hubble Space Telescope} \citep{Cool95}, 
along with a population of "nonflickerers" \citep{Cool98} which are understood 
to be young helium white dwarfs with C/O white dwarf companions \citep{Hansen03}.

In the past few years,  substantial progress has been made in
optical identification of {\it Hubble Space Telescope} counterparts to
 {\it Chandra} X-ray sources in several GCs.
Valuable information was obtained for populations of CVs, 
chromospherically active binaries and quiescent low-mass X-ray binaries (qLMXBs) 
\citep{Grindlay01a, Pooley02a, Edmonds03a, Bassa04, Heinke03a}.
For the first time we can compare populations of such binaries in
globular clusters (GCs) and in the Galactic field, and infer their rates of
formation and population characteristics.
In particular, 22 CVs have now been identified in 47 Tuc, allowing 
identification of several differences between typical CVs in globular 
clusters  and CVs in the Galactic field.  
These differences include relatively high X-ray luminosities 
compared to field systems \citep{Verbunt97}; 
a lack of novae, and of the steady, bright blue accretion discs signifying 
novalike CVs, in GCs \citep{Shara95}; 
 relatively low frequencies of  dwarf nova outbursts (DNOs), the typical identifiers of CVs in the Galactic disc
\citep{Shara96}; 
and a higher ratio of X-ray to optical flux than in most field CVs \citep{Edmonds03b}.
These differences produce puzzles: the lack of novae, novalikes, and DNO suggests very 
low mass transfer rates, while the high X-ray luminosities indicate moderate mass 
transfer rates.  The X-ray to optical flux ratio suggests the CVs are DNe, but  
the lack of DNO argues against this.
It was suggested that CV discs in GCs are more stable due to a combination of 
low mass transfer rates and moderately strong white dwarf magnetic moments \citep{Dobrotka05}. 
This hints that the evolutionary paths of CVs in GCs and in the field are different.
Comparisons of the numbers of CVs in clusters of different central densities also supports the idea  
that CVs are produced through dynamical interactions \citep{Pooley03}, though there is an 
indication that CV production may depend more weakly on density than the production of 
low-mass X-ray binaries containing neutron stars \citep{Heinke03a}.

This is the first of two papers where we summarize results of our studies on compact binary formation in GCs, 
some preliminary results of which were reported in \citet{Ivanova04a,Ivanova04b,Ivanova04c}.
In this paper we focus on the formation of compact binaries with a white dwarf, and in the second
paper (Paper II) we will describe dynamical formation and evolution of binaries with a NS companion.
We explore a large spectrum of globular cluster models, where for the first time we take into account
(i) the mechanism of binary formation through physical collisions using results from 
smoothed-particle hydrodynamics (SPH) and
(ii) the effect of metallicity on the formation and subsequent evolution of close binaries.
In Section~2 we provide a complete review of the physical processes  
of formation and destruction of mass-transferring WD binaries.
In Section~3 we outline the methods and assumptions.
The major formation channels, and population characteristics for CVs 
and AM~CVn systems (double WD systems where one WD experiences Roche lobe 
overflow) in different clusters are presented and discussed in Section~4.
We conclude in the last section by addressing the connection between our 
results and the observations.

\section{Mass-transferring WD-binaries in a dense cluster}

There are several ways to destroy a primordial binary in a globular cluster.
For instance, in a dense region a soft binary will very likely be  ``ionized'' (destroyed)
as a result of a dynamical encounter. 
A hard binary, in contrast, can be destroyed through a physical 
collision during the encounter. 
The probability of such an outcome
increases strongly as the binary becomes harder \citep{Fregeau04}.
In addition to dynamical processes, a primordial binary can be destroyed 
through an evolutionary merger or following a SN explosion. 
Overall, even if a cluster initially had
100\% of its stars in binaries initially,
the binary fraction at an age of 10-14 Gyr will typically be as low as $10\%$ 
\citep{Ivanova05}.

To understand the evolution of a primordial binary in a dense environment and 
the probability of a binary becoming a CV, two steps are required: (i) compare
the evolutionary time-scales with the time-scale of dynamical encounters;
(ii) analyze what is the consequence of an encounter (this depends strongly on the 
hardness of the binary).

The time-scale for a binary to undergo a strong encounter with another
single star (the collision time) can be estimated as 
$\tau_{\rm  coll}=(n\Sigma v_\infty)^{-1}$.
Here $\Sigma$ is the cross section for an encounter between two objects, 
of masses $m_i$ and $m_J$, with relative velocity at infinity $v_\infty$ 
and is given as
\begin{equation}
\Sigma = \pi d_{max}^2 (1+v_{p}^2 / v_{\infty}^2)\ ,
\end{equation}
where $d_{max}$ is the maximum distance of closest approach that defines a significant encounter
and $v_{p}^2 = 2G (m_i + m_j)/d_{max}$ is the velocity at pericenter. 
Assuming that a strong encounter occurs when the distance of closest  approach
is a few times the binary separation $a$, $d_{max}\le k a$ with $k\simeq 2$, we obtain

\begin{eqnarray}
\label{tcoll_pd}
\tau_{\rm coll} = 3.4 \times 10^{13} \ {\rm yr} \ \  k^{-2} P_{\rm  d}^{-4/3} M_{\rm tot}^{-2/3} n_5^{-1}
v_{10}^{-1} \times \\ \nonumber
\left( 1+913 {\frac { (M_{\rm tot} + \langle M\rangle )} {k P_{\rm d}^{2/3} M_{\rm  tot}^{1/3} v_{10}^2}}\right) ^{-1}
\end{eqnarray}
 
\noindent Here $P_{\rm d}$ is the binary period in days, 
$M_{\rm tot}$ is the total binary mass in  $M_\odot$,
$\langle M\rangle$ is the mass of an average single star in  $M_\odot$,
$v_{10}=v_{\infty}/(10\,{\rm km/s})$ and $n_5=n/(10^5\,{\rm pc}^{-3})$, where $n$ is the stellar number  density.

The hardness of a binary system, $\eta$, is defined as  
\begin{equation}
\label{eta_def}
\eta = {\frac {G m_1 m_2} {a \sigma^2 \langle m\rangle }}\ ,
\end{equation}
\noindent where $a$ is the binary separation, $\sigma$ is the central velocity dispersion, 
$m_1$ and $m_2$ are the masses of the binary components,
and $\langle  m\rangle$ is the average mass of a single star.
Binaries that have $\eta < 1$ are termed soft, and those with $\eta > 1$ are termed hard.

\subsection{Primordial CVs and AM~CVns.}

\begin{figure}
  \includegraphics[height=.35\textheight]{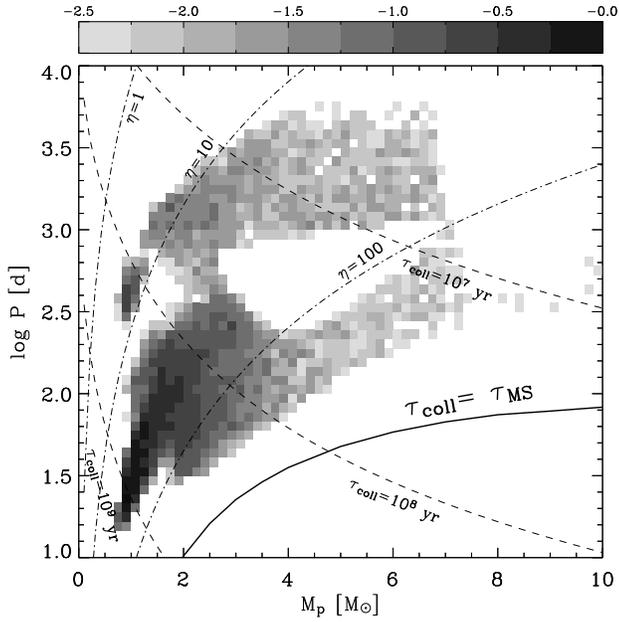}
  \caption{ Distribution density of CV progenitors (initial masses of primary stars $M_{\rm p}$ and 
binary periods $P$) for non-eccentric binaries in the Galactic field, 
with Z=0.001. The total normalization of CV progenitors is scaled to unity, the grey color
shows $\log_{10}$ of the normalized distribution density.  
The thick solid line indicates the binary period where the collision time 
of the binary is equal to the main sequence lifetime of the primary 
(using a core number density $n=10^5\,{\rm pc}^{-3}$, a central velocity 
dispersion 10 km/s and an average object mass of $0.5 \,M_\odot$).
Dash-dotted lines are lines of constant binary hardness and dashed lines 
are lines of constant collision time.
}
\label{cv_field}
\end{figure}

The typical formation scenario for CVs in the field (low density environment) 
usually involves common envelope (CE) evolution.
In Fig.~\ref{cv_field} we show parameters of primordial non-eccentric  binaries that
successfully become CVs. 
To obtain this parameter space, we used the binary population synthesis code 
{\tt StarTrack} \citep{Bel02,Bel05b}\footnote{For the calculations in this study, we have used the StarTrack
     code prior to the latest release \citep{Bel05b}.
     However, the most important updates and revisions of input physics, in particular the ones
     important for evolution of binaries with white dwarfs, were
     already incorporated in the version we have used.}.
We evolved $5\times10^5$ binaries considering specifically that region of primordial binaries
which, according to preliminary lower resolution runs, 
leads to CV formation. Our primary stars have masses between 0.5 $M_\odot$ and 10 $M_\odot$,
the secondaries have masses according to a flat mass ratio distribution with initial periods distributed flatly between 1 and $10^4$ days.
For demonstration purposes in Fig.~\ref{cv_field}, we use initially circular orbits,
because the parameters leading to different formation channels can be more clearly distinguished.
For our actual cluster simulations we use eccentric binaries;
in comparison to Fig.~\ref{cv_field}, eccentric primordial binaries can have higher initial periods and still produce CVs. 
Progenitors of CVs with a MS donor are located in the left bottom corner, with 
$M_{\rm p}\la 4\,M_\odot$ and $\log P\la 2.5$.
In other cases the donor star is a red giant (RG) or a (subgiant) star in the Hertzsprung gap.
For primordial binaries located in a small but dense area at the left middle part of Fig.~ \ref{cv_field},
$\log P\sim 2.7$ and $M_{\rm p}\sim 1\,M_\odot$, a CE does not occur.
We note that the lifetime of a binary in the CV stage with a RG donor 
is about 1000 times shorter than in the case of a MS donor.
 
\begin{figure}
  \includegraphics[height=.35\textheight]{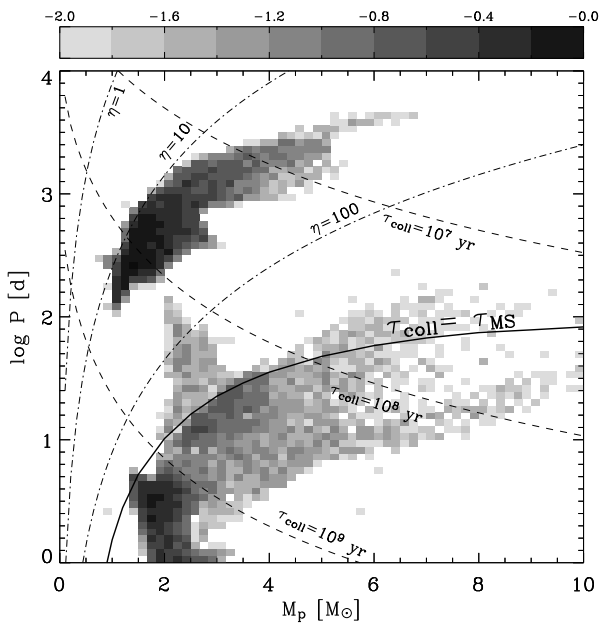}
  \caption{Distribution density of AM~CVn progenitors (initial masses of primary stars $M_{\rm p}$ and 
binary periods $P$) for  non-eccentric binaries in the Galactic field, Z=0.001. 
Notation as for Fig.~\ref{cv_field}.
}
\label{amcv_field}
\end{figure}

In the core of a GC with core density $\rho_{\rm c} \sim 10^5$ pc$^{-3}$, 
a binary with an initial period typical of a CV progenitor 
will experience a dynamical encounter before its primary 
leaves the MS (see Fig.~\ref{cv_field}, where all CV progenitors lie above 
the line indicating equality of $\tau_{coll}$ and $\tau_{MS}$).
The unaltered primordial channel for CV formation is therefore likely to succeed 
only for binaries that enter the dense cluster core after their CE event; 
the post-CE binary is compact enough to avoid an encounter. 
The contribution of the primordial channel depends therefore on the time --
before or after the moment of CE -- when primordial CV binaries will segregate into the central dense core.
In more detail, an average initial binary in the GC is $\sim 0.7 M\odot$, which is significantly
smaller than the pre-CE mass of a primordial CV binary (see Fig.~\ref{cv_field}).
Post-CE primordial CV binaries are also heavier than typical binaries in the halo (for which the average binary mass is $\sim 0.4 M\odot$).
In both cases, primordial CV binaries, as heavier objects, will tend
to sink toward the cluster core 
on the cluster half-mass relaxation time.

The situation is similar for the formation of AM~CVn systems from 
primordial binaries (see Fig.~\ref{amcv_field}).
In this case, the main formation channel requires the occurrence
of two CE events (see also \cite{Bel05a}),
and the primordial binary is expected to be even wider. 
However, the second channel, with two stable MT stages (at the 
start of the RG stage of the primary, and when the secondary
becomes a helium giant), is provided by relatively compact progenitor 
binaries. These binaries are expected to 
evolve in the same way in a GC as in the field.

\subsection{Dynamical formation of CVs}

A binary consisting of a MS star and a WD can be formed via 
several kinds of dynamical encounters:
via an exchange interaction, via a tidal capture (TC) of a MS by a WD, or 
via physical collisions between a red giant (RG) and a MS star.
A fraction of these dynamically formed MS-WD binary systems 
will start MT and become a CV.
In this section we examine in detail the possible channels for CV creation.

\begin{figure}
  \includegraphics[height=.35\textheight]{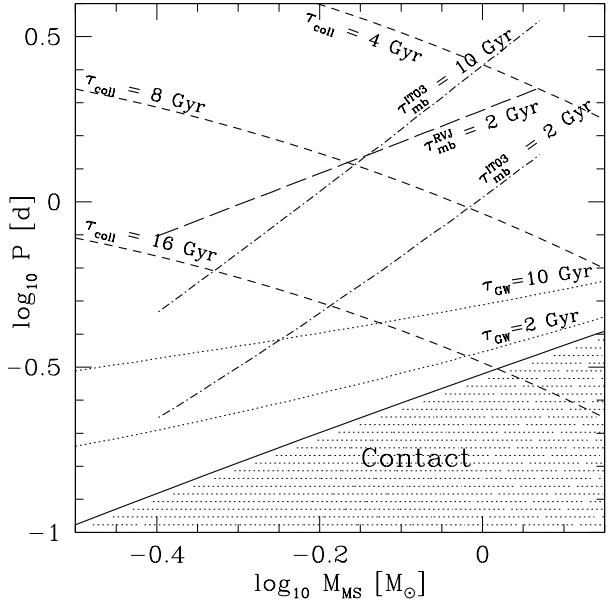}
  \caption{The fate of non-eccentric MS-WD binaries 
produced by, e.g., dynamical encounters, 
where the primary is a WD of 0.6 $M_\odot$.
$P$ is the post-encounter (or post-CE) orbital period and $M_{\rm MS}$ is the mass 
of a MS secondary.  The short-dashed lines show the binary periods for 
constant collision times and the dotted lines delineate the binaries that 
will shrink within 2 and 10 Gyr due to gravitational wave emission.
The long-dashed line indicates the upper period limit for binaries that 
will begin MT within 2~Gyr
with the RVJ MB prescription, while the dash-dotted lines indicate 
those that will begin MT within 2 and 10 Gyr with IT03 MB.
Below the solid line the binary is in contact.
\label{mswd-nonecc}
}
\end{figure}

\begin{figure}
  \includegraphics[height=.35\textheight]{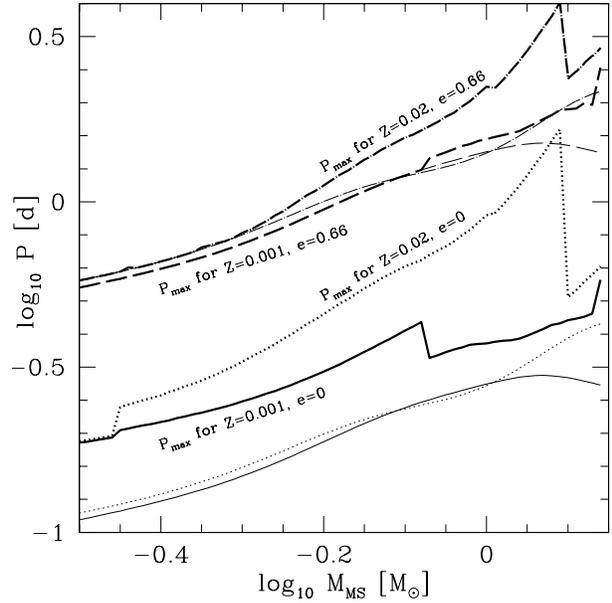}
  \caption{The fate of post-encounter MS-WD binaries 
where the primary is a WD of 0.6 $M_\odot$, for post-exchange 
eccentricities 0 and 0.66.  $P$ is the post-encounter orbital period 
and $M_{\rm MS}$ is the mass of the MS secondary.  Thick lines delineate 
the maximum periods for binaries which will begin MT within 2 Gyr. 
Thin lines of the same type show the period at which that binary will 
begin MT.  
\label{mswd-ecc}
}
\end{figure}

The main angular momentum losses in  a close MS-WD binary occur via 
magnetic braking (MB)
and  gravitational wave (GW) emission, both of which lead to orbital decay.
In eccentric binaries, the binary orbital separation will be affected 
by tides, and the post-circularized periastron is larger 
than the pre-circularized periastron (unless tidal synchronization 
is significant). 
In Fig.~\ref{mswd-nonecc} we show the maximum initial periods (at the moment 
of the binary formation) of a non-eccentric MS-WD binary that can start MT 
within 2 Gyr, and within 10 Gyr, due only to GW or only to MB 
(for illustrative purposes, we show time-scales for two prescriptions
of magnetic braking, one is standard MB according to \cite{RVJ} (RVJ) and
the second is the MB based on dipole-field model according to \cite{Ivanova03} (IT03)). 
A maximum initial period such that a binary is able to start MT without
having any other encounters is only $\sim$2 days.
On Fig.~\ref{mswd-ecc} we again show the maximum initial periods of binaries 
that may start MT, but now including all angular momentum losses (GW, MB and tides),  
and compare the cases of non-eccentric and eccentric binaries.  On this figure 
we also show the difference in maximum initial period between metal-poor 
and metal-rich GCs. In metal-poor clusters only stars with $M\la 0.85\,M_\odot$
have developed outer convective zones, allowing MB and convective tides to 
operate \citep{Ivanova06}.
This effect can potentially be dramatic; for instance, among non-eccentric 
binaries with a MS star of $1\,M_\odot$, the range of 
post-exchange periods that leads to CV formation is a factor of 6 
larger if the donor has Z=0.02, compared to Z=0.001. 
For eccentric binaries, this ratio is higher, as 
tidal circularization via radiative damping will
reduce binary eccentricity (and therefore increase the periastron) 
more effectively than GW can shrink the binary orbit.

\begin{figure}
  \includegraphics[height=.35\textheight]{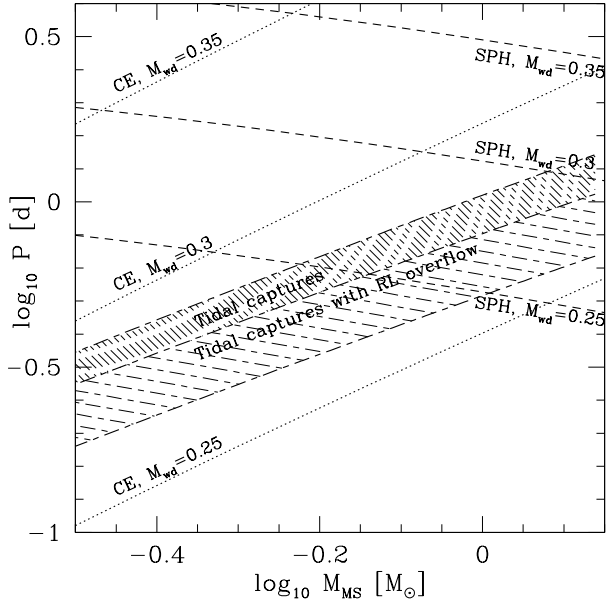}
\caption{Formation of WD-MS binaries via physical collisions and tidal captures.
The hatched area shows binaries formed via TC with a 0.6 $M_\odot$ WD.  
In the dense hatched area, the MS star did not overflow its Roche lobe 
at the minimum approach during the TC. The dashed lines show binaries 
formed via physical collisions of a MS star and a 0.8 $M_\odot$ RG,
 for different core masses, using parameterized results of SPH simulations
(for illustrative purposes, we show only the case of 
the impact parameter to be 0.54 of the RG radius with corresponding 
post-collisional eccentricity of 0.7). 
The dotted lines show binaries formed via physical collisions of MS star and a 0.8 $M_\odot$ RG,
 for different core masses, assuming  common envelope approach ($\alpha_{\rm CE} =\lambda=1$).
\label{mswd-formation}
}
\end{figure}

A circular binary is most likely to be formed via tidal capture (TC),
where a post-capture circularization is assumed. 
Using the approach described in \cite{Zwart_TC_93}, we can estimate   
the post-capture binary parameters for a MS-WD binary 
(see Fig.~\ref{mswd-formation}, where WD mass is assumed to be 0.6 $M_\odot$).  
The upper limit here corresponds to the closest approach at which tidal interactions
are strong enough to make a bound system, and the lower limit corresponds
to the closest approach at which the MS star overfills its Roche lobe by 1/3.
We note that the parameter space for tidally captured binaries 
where the MS star does not overfill its Roche lobe at the closest approach 
is very small (see Fig.~\ref{mswd-nonecc}).
We note that this is an optimistic estimate, as the captured star can also be destroyed 
during the chaotic phase of the tidal energy damping \citep{Mardling95_chaos2}.
Most tidally captured binaries can be brought to contact by MB before 
either the next encounter occurs, or the MS star evolves away from the MS. 

An eccentric binary can be formed via an exchange encounter or a 
physical collision; eccentricity can also be increased via the cumulative effect of fly-by encounters.
For binaries formed through MS-RG collisions, the post-exchange  binary 
separation $a_{\rm f}$ as well as post-exchange eccentricity
$e_{\rm f}$ depends on the closest approach $p$ \citep{Lombardi_2006}
and can be estimated using results of SPH simulations.
These simulations were done for physical collisions of a NS and a RG,
and therefore are not straightforwardly applicable for the physical collisions
of a MS star and RG. We therefore study how strongly 
the choice of the treatment can affect the final results.
We consider the two following prescriptions:

\begin{itemize}

\item {Using a common-envelope (CE) prescription:

\begin{equation}
{\frac{(M_{\rm rg}+M_{\rm ms})v_{\infty}^2}{2}}  + \alpha_{\rm CE} {\frac{GM_{\rm wd}M_{\rm ms}} {2 a_{\rm f}}} = 
{\frac {G M_{\rm rg} (M_{\rm rg}-M_{\rm wd} )} {\lambda R_{\rm rg}}}
\label{af_sph}
\end{equation}

\noindent Here $M_{\rm rg}$, $M_{\rm ms}$ and $M_{\rm wd}$ are the 
masses of the RG, MS star, and RG core that will become a WD, in $M_\odot$; 
 $R_{\rm RG}$ is the RG radius; 
$\alpha_{\rm CE}$ is the CE efficiency parameter; 
and $\lambda$ is the CE parameter that connects a star's binding energy
 with its parameterized form.
We assume that after a common envelope event the binary is not eccentric.}

\item {Using parameterized results from SPH simulations:

\begin{equation}
e_{\rm f} = 0.88-{\frac {p} {3 R_{RG}}}
\end{equation}
\begin{equation}
a_{\rm f} = {\frac {p} {3.3 (1-e_{\rm f}^2)}}
\label{af_ce}
\end{equation}

\noindent As the parameterized SPH simulations were done for a limited set of mass ratios,
we also check the energy balance. When we consider 
the case of the second treatment, we choose 
the minimum binary separation from  eq.~(\ref{af_sph}) and (\ref{af_ce}),
as at small masses the extrapolated prescription from SPH simulations can lead
to the formation of binaries with artificial energy creation.
Also, in the case when a MS star at the pericenter overfills its Roche lobe,
we destroy the MS star instead of forming a binary. This is consistent with
the results of SPH simulations for physical collisions of a RG  and a MS star 
(J.~Lombardi 2005, priv. communication).
}
\end{itemize}

In Fig.~\ref{mswd-formation} we also show possible binary periods for binaries 
formed via physical collisions with a red giant. Note that it is hard to 
form a relatively close MS-WD binary (one that is able to start MT within 
a few Gyr) with a WD more massive than 0.3 $M_\odot$ via either prescription. 
Also, in binaries with the mass ratio $\ga 3$, Roche lobe overflow leads
to delayed dynamical instability and a binary merger.
This limits the MS star mass to $\la 0.9\,M_\odot$.
Therefore, the CV progenitors from the channel of physical collisions of 
RGs and MS  stars are expected to initially have rather low mass WD 
accretors, and donor star masses $\la 0.9\,M_\odot$. 
Therefore, metallicity variations should not affect this channel strongly.
The evolutionary stage during He core burning lasts almost 
the same time as the RG branch, however, He core stars of $\la 2\,M_\odot$ are 
a few times more compact than at the
end of the RG branch and have a larger core than during RG evolution.    
Therefore a collision between a He core burning star and a MS star also favors
the formation of a WD-MS binary that is close enough to become a CV.
This channel mainly provides binaries with a WD mass at the start of  
accretion of about 0.5 $M_\odot$ 
(just a bit above the core mass at the time of He core flash).

On the other hand, there are not many single WDs of such small masses 
present in a GC core.  
A WD with mass $\la 0.3\,M_\odot$ cannot (yet) be formed in single star evolution --
it must evolve via a CE event or a physical collision. 
A binary containing such a WD is very hard and has $\tau_{\rm coll}\ge 10$~Gyr.
If an encounter occurs, it is more likely to result in a merger 
rather than an exchange.  We therefore expect that most CVs with a 
low mass WD companion will be formed either through a CE event 
(in a primordial binary or in a dynamically formed binary with 
$P\sim 10-100$ days), or as a result of a physical collision, but not via direct exchange encounter.

A typical binary formed via an exchange encounter has $e\approx 0.7$.
In order to become a CV within 2 Gyr (or before the next encounter), 
it should have a post-encounter period of a few days 
(see also Fig.~\ref{mswd-ecc}).
According to energy conservation during an exchange encounter \citep{Heggie96},
and assuming that during an exchange encounter the less massive companion 
is replaced by the more massive intruding star, the post-encounter binary 
separation will be larger than pre-encounter.  The domain of 
pre-encounter binaries that will be able to form a CV-progenitor binary
via only exchange encounter is therefore limited to very short period 
binaries (with correspondingly long collision times), and these binaries 
are very likely to experience a physical collision rather than an exchange 
\citep{Fregeau04}.

Let us consider the possibilities for an initially wider dynamically 
formed binary than shown on Fig.~\ref{mswd-ecc}
to evolve toward MT. For definiteness, we consider a binary consisting 
of a 1 $M_\odot$ MS star and a 0.6 $M_\odot$ WD with an initial period of 
10 days. 
There are two kinds of post-formation dynamical effects that can happen during
fly-by encounters: (i) binary hardening;
(ii) eccentricity pumping.
Even if each hardening encounter could reduce the orbital separation 
by as much as $50\%$, the hardening of this binary from 10 days to 1 day  
(at this period MB starts to be efficient) will take about 20 Gyr.
In the case of eccentricity pumping (assuming no binary energy change), 
the mean time between successive collisions stays at $\tau_{\rm coll}\le1$~Gyr 
and therefore a binary can experience many encounters. 
If the acquired eccentricity $e\ge0.95$, the binary can shrink through 
GW emission even if its initial period is larger than 10 days.
The last possibility for such a wide dynamically formed binary to become a CV
is a CE event that happens in a post-exchange MS-MS binary.

\subsection{Dynamical formation of AM~CVns}

\begin{figure}
  \includegraphics[height=.35\textheight]{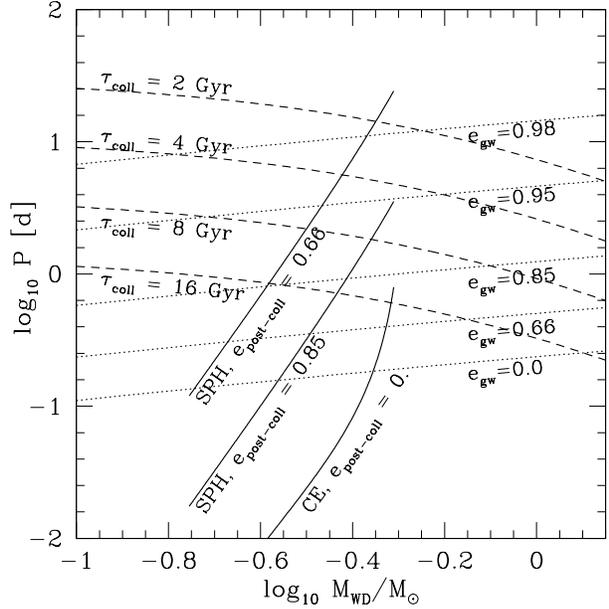}
  \caption{The fate of post-encounter WD-WD binaries where the 
primary is a WD of 0.6 $M_\odot$.  $P$ is the post-encounter orbital 
period and $M_{\rm WD}$ is the mass of the WD secondary.  
The dashed lines show the binary periods for constant collision times and
the dotted lines delineate the binaries that will begin MT within 2 Gyr 
due to GW emission for different post-encounter eccentricities.
The solid lines show binaries of different eccentricities 
that can be formed through a collision between a WD of 0.6 $M_\odot$
and a RG of 0.8 $M_\odot$ ( $\alpha_{\rm CE} \lambda=1$).  
\label{wdwd}
}
\end{figure}

For the evolution of WD-WD binaries, we adopt that only GW are important 
as a mechanism of angular momentum loss and neglect the possibility of tidal heating.
The maximum possible periods for different post-encounter eccentricities are
shown on Fig.\ref{wdwd}. 

Let us first examine a WD-WD binary formation via direct exchange.
Again, as in the case of MS-WD binaries, a typical eccentricity 
 is $e\sim0.7$ and the separation is comparable to the pre-exchange separation.
The collision time for both pre-encounter and post-exchange binaries is so long
that both binary hardening and exchanges are very rare events.
The main difference with MS-WD binaries is that post-exchange WD-WD binary periods
that will allow a binary to evolve to mass transfer (MT) are several times smaller 
for the same eccentricities. Therefore, the exchange channel producing a 
post-exchange binary consisting of two WDs seems to be very unlikely.

A more important channel
seems to be the case of an exchange encounter that leads to the
formation of a MS-WD binary.
If the MS star is massive enough to become a RG during the cluster lifetime,
such a binary can evolve through CE and form a close WD-WD binary.

The second important channel is again a physical collision, involving
a single WD with a RG (see Fig.~\ref{wdwd},  where we show 
possible outcomes of a such a collision). We note that both treatments 
(parameterized SPH results and CE prescription) lead to the formation of
 WD-WD binaries that are roughly equally likely to start the MT.

We therefore expect that only a post-CE system can become an AM~CVn, 
where the post-CE system could be from a primordial binary, a 
post-collision binary, or a dynamically formed binary. 

\section{Methods and assumptions}

For our numerical simulations of globular clusters we use a Monte Carlo approach described in detail in 
\citet{Ivanova05}.
The method couples the binary population synthesis code
{\tt StarTrack} \citep{Bel02,Bel05b}, 
a simple model for the cluster, and a small 
$N$-body integrator for accurate treatment of all relevant dynamical 
interaction processes \citep[{\tt FewBody}, ][]{Fregeau04}.
The main update of the code is the treatment of physical collisions with a RG,
for which we now use the parameterized results of SPH simulations from 
\citet{Lombardi_2006} as described in \S 2.2.
In our code we keep a complete record of all events that happen to any cluster star,
dynamical (like  collisions, tidal captures and exchanges, as well as 
changes of the binary eccentricity or the binary separation after a fly-by encounter),
or evolutionary (like common envelope events, mass transfers, or SN explosions).
This helps to analyze the final populations and understand what factors played the most significant role in their formation.

The ``standard'' cluster model in our simulations has initially $N=10^6$ stars 
and initial binary fraction of 100\%. 
The distribution of initial binary periods is constant
in the logarithm between contact and $10^7$~d and the eccentricities are distributed thermally.
We want to stress here that about 2/3 of these binaries are soft initially (the 
binary fraction provided by only hard binaries gives an initial binary fraction of about $20\%$ if the 1-D velocity dispersion is 10 km/s) 
and most very tight binaries are destroyed through  evolutionary mergers.
Our initial binary fraction is therefore comparable to the initial binary 
fractions that are usually used in $N$-body codes,
where it is assumed for simplicity that very soft binaries
will not live long as binaries and will only slow down the simulations.  
For more detailed discussion on the choice of the primordial binary fraction, see \cite{Ivanova05}. 

For single stars and primaries we adopted the broken power law initial mass function (IMF) of \citet{Kroupa02} 
and a flat mass-ratio distribution for secondaries. The initial
core mass is 5\% of the cluster mass and, assuming initial mass segregation,
an average object in the core is about twice as massive as 
an average cluster star.
At the age of 11 Gyr the mass of such a cluster in our simulations is $\sim 2\times 10^5\,M_\odot$
and is comparable to the mass of typical globular clusters in our Galaxy.

We adopt a core number density  $n_{\rm c}=10^5 \ {\rm pc}^{-3}$ (this corresponds to  
$\rho_{\rm c}\approx 10^{4.7}\,M_\odot \ {\rm pc}^{-3}$ at the ages of 7-14 Gyr), 
a half-mass relaxation time $t_{\rm rh}=1$ Gyr and a metallicity  $Z=0.001$.
The characteristic velocities are taken as for a King model $W_0=7$ for the cluster of this mass.
We take a one-dimensional velocity dispersion 
$\sigma_1=10$ km/s and an 
escape velocity from the cluster $v_{\rm esc}=40$ km/s.
If, after an interaction or SN explosion, an object in the core acquires a velocity
higher than the recoil velocity $v_{\rm rec}=30$ km/s, an object is moved from the core to the halo.
The ejection velocity for objects in the halo is $v_{\rm ej,h}=28$ km/s.

In  addition to the ``standard'' model 
we also considered cluster models with the following modifications:
\begin{itemize}
\item a metal-rich cluster with $Z=0.02$ (``metal-rich'');
\item central density $n_{\rm c}=10^4 \ {\rm pc}^{-3}$  (``med-dens'') or  $n_{\rm c}=10^3 \ {\rm pc}^{-3}$  (``low-dens'') 
\item initial binary fraction 50\%  (``BF05'');
\item RVJ magnetic braking (``fast MB'');
\item treatment of physical collision using a CE prescription (``CE coll'');
\item 47~Tuc-type cluster, characterized by  a higher density
$\rho_{\rm c}=10^{5.2}\,M_\odot \ {\rm pc}^{-3}$, higher metallicity $Z=0.0035$,
$\sigma_1=11.5$ km/s, $v_{\rm esc}=57$ (with the recoil velocity 
of 52 km/s and $v_{\rm ej,h}=24$ km/s) and $t_{\rm rh}=3$ Gyr (``47 Tuc'').
\end{itemize}

``47 Tuc'' model describes the GC where currently the largest
CV population is identified. 
In order to find a better match with the observations, we examined
several variations of this model.
In particular, we considered a model with an initial binary population of 50\% (``47 Tuc+BF05''),
and a model with an initial binary population is 50\%, and the initial core has a smaller mass - 2\%, 
reflecting the effect of a longer half-mass relaxation time on the initial population (``47 Tuc+SCBF05'').
We also examined the sensitivity of the final CV production 
to the CE efficiency parameter, considering the 
case with $\alpha_{\rm CE}\lambda=0.1$ (``47 Tuc+$\alpha_{\rm CE}\lambda$'').

In order to estimate the effects of dynamics on the population we also ran the same population
as in our ``standard'' model (Z=0.001), but without dynamics (``non-dyn'').
In order to compare to a field population, we considered the population of stars
with solar metallicity Z=0.02 and with different times of star formation, assuming flat
star formation rate through last 10 Gyrs (``field''). 
In ``non-dyn'' model all stars are formed at the zero age, like in GCs.

\begin{figure*}
  \includegraphics[scale=0.7]{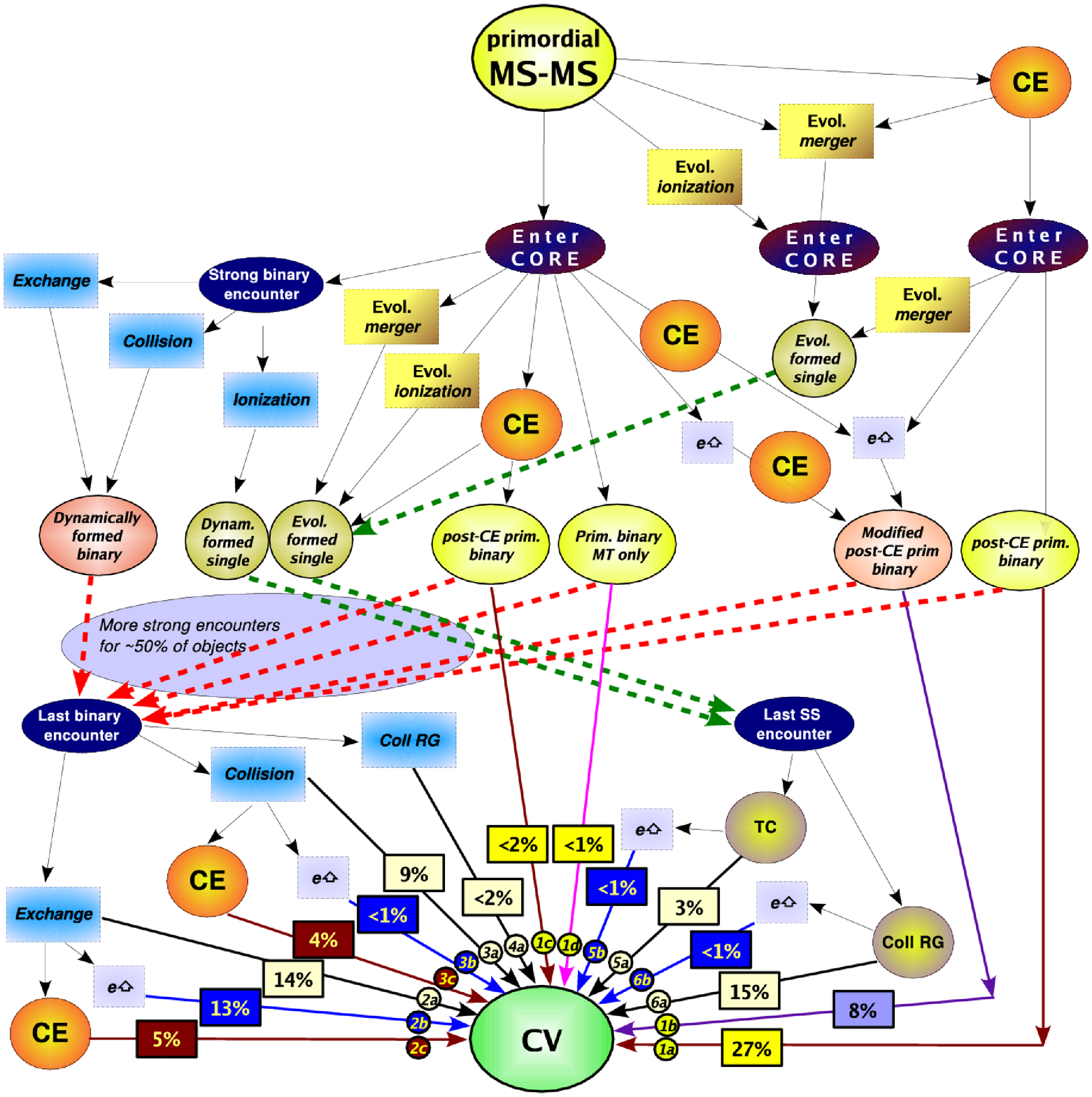}
  \caption{Main formation channels of all CVs that are present in a ``standard'' cluster at the age of 10 Gyr. 
Notations: numbers in squares represent contributions from different channels, small circles contain the 
labels for each channel. 
``CE'' - common envelope, ``Evol. merger'' - evolutionary merger, ``Evol. ionization'' - destruction of the binary via SN explosion,
``e$\Uparrow$'' -- increase of the eccentricity during fly-by encounters, ``Strong binary encounter'' -- three- or four-body
encounter with an outcome as exchange, binary destruction (``ionization'') or  physical collision (``Collision'');
``TC'' -- tidal capture, ``Coll RG'' -- physical collision with a RG.
}
\label{cv-scen-blue}
\end{figure*}

\section{Numerical results}

\subsection{Formation channels of CVs}

\subsubsection{Main formation channels in the ``standard'' model}

\label{form-chan}
 
In Fig.~\ref{cv-scen-blue} we show the formation channels for all CVs 
that are present in a typical cluster (our ``standard'' model) at the age of 10 Gyr.
Most of these CVs are too dim to be detected, and we consider separately 
the population of CVs that can be detectable according to present 
observational limits, considered specifically for the globular cluster 47 Tuc.
For the limiting X-ray luminosity, we take $L_{\rm x}\ga 3\cdot 10^{30}$ ergs s$^{-1}$ \citep{Grindlay01a},  
and the limiting bolometric luminosity of the donor $L_{\rm d}\ga 0.06 L_{\odot}$, 
set by the limiting magnitude of HST in the cluster core \citep{Edmonds03a}.

We label channels in the following way: the first character indicates 
the last major dynamical event affecting the 
binary before it becomes a CV in the core, as follows:
(1) entering the core (primordial binary);
 (2) companion exchange during a binary encounter; 
(3) merger during a binary encounter; 
(4) physical collision with a RG during a binary encounter that 
resulted in a tight binary formation with a RG core as a companion; 
(5) tidal capture;
(6) physical collision of a single MS star with a single RG.  
The second character indicates a sub-channel by which the binary was 
modified after the last major dynamical event occurred:
(a) stands for all sub-channels where no strong evolutionary or dynamical 
event occurred; (b) eccentricity of the formed binary was increased via 
 binary-binary encounters; (c) common envelope occurred; (d) a previous MT 
episode played the most important role in the orbital decay.

The {\it primordial channel} ({\bf channel 1}) -- provides  $37\%$ 
of all CVs that are present in the cluster core ( 42\% of detectable CVs).
We call this channel primordial as the binary keeps both its initial companions,
and no mergers ever occurred to either of them.
Only 3/4  of CVs formed via this channel are ``purely'' primordial in the sense that they
did not experience a significant dynamical encounter throughout their life ({\bf 1a}, 
see also Fig.~\ref{cv-scen-blue});
most of these ``purely'' primordial CVs  evolved via CE before they entered the core.
As was predicted in \S~2.1, very few CVs come from the channel where CE occurred
after a binary entered the core ({\bf 1c}).
1/5 of all primordial CVs  would not evolve via CE or start a MT unless 
their eccentricity was increased via fly-by encounters ({\bf 1b}). 
A small fraction of primordial CVs evolved without a CE but with only a 
MT episode on to a MS star ({\bf 1d}), as was described in \S~2.1.

The binary encounters ({\bf channel 2, 3 and 4}) are responsible for the 
formation of $46\%$ of all CVs, and the same fraction of detectable CVs. 
In most cases the binary that participated in the binary encounter
was not a primordial binary, but a dynamically formed binary.  
In more than half of cases, a future accretor had been a companion
in at least 3 different binaries before it acquired its final donor.

The most effective path is the {\it binary exchange channel} 
({\bf channel 2}) -- it provides 32\% of all CVs.
Within this channel, $\sim 40\%$ of post-exchange binaries evolved toward the MT 
without further significant dynamical or evolutionary events ({\bf 2a}),
in 20\% of them CE occured ({\bf 2c}) and in 40\% of them the MT started 
as a result of the
eccentricity pumping during subsequent fly-by encounters ({\bf 2b}).
This is the most efficient channel for eccentricity pumping.

Exchange encounters that lead to CV formation typically occur between 
the following participants:
(i) a single, relatively heavy WD  (about  $0.7-1.4\,M_\odot$) 
and a MS-MS binary of  total mass $\la 1\,M_\odot$;
(ii) a single, relatively massive MS star (about turn-off mass) 
and a MS-WD or WD-WD binary.
In the latter case, CE often follows the exchange encounter.
The number of successful encounters between MS star and WD-WD binary 
is relatively small, and no successful four-body encounter occurred.
Nearly all binaries that proceed via sub-channels 2a or 2b are 
WD-MS binaries, and all binaries in 
sub-channel 2c are MS-MS binaries after the last strong binary encounter.
A post-exchange binary typically has a heavier WD than a primordial (post-CE) binary has.

A further $13\%$ of CVs are formed in binaries that experienced a 
physical collision  during the last three- or four-body encounter -- 
{\it binary collisional channel} ({\bf channel 3}), while in 1\% of 
cases a physical collision with a RG occurred during the encounter and a binary
with the stripped RG core was formed ({\bf channel 4}).
In the evolution of post-collisional binaries the eccentricity change 
plays a smaller role compared to post-exchange binaries; 
MT is started due to the evolutionary angular momentum losses.

The {\it tidal capture channel} ({\bf channel 5}) contributed very little in our standard model.
When we looked at all CVs that were formed via TC over all GC ages, we find 
that a typical WD that captured a MS star is $\sim 1.0\pm0.2\,M_\odot$.
In our simulation we allowed a star to overfill its Roche lobe radius by up 
to 1/3  during the tidal capture encounter and survive.
If all encounters where a MS star overfills its Roche lobe 
lead to the stars' merger, then the contribution of tidal
captures would be even smaller.

Finally, the {\it channel of physical collision with RGs} ({\bf channel 6}) 
provides $15\%$ of all CVs but much smaller fraction of detectable  CVs.
Eccentricity pumping played a very small role in both TC and physical collision channels.
Typical participants of a  successful physical collision (leading to CV formation) 
are a  MS star of $0.3-0.9~M_\odot$ and 
a RG of about $1-1.7\,M_\odot$  with a core around $0.3\,M_\odot$ or 
a He core burning giant with a core mass around $0.5\,M_\odot$. 
CVs formed by this channel are similar to post-CE CVs from primordial binaries.
We also compared the results of CVs productions in our large model with $10^6$ stars and in the model
with three times less stars. We noted that, with the increase of the resolution,
the total number and the number of detectable CVs per unit of the core mass is slowly decreasing.
Branching ratios between sub-channels within a channel can vary slightly, but 
an overall picture is the same.

We outline our main findings:
\begin{itemize}
\item{Only $\sim 25\%$ of CVs were formed in binaries that would become CVs in the field.}
\item{In $\sim 20\%$ of CVs the main reason for a binary to become a CV were fly-by encounters. 
These CVs cannot be predicted in simulations where only strong encounters are taken into account.}
\item{In $\sim 15\%$ of CVs, the WD was formed during dynamical removal 
of the RG envelope.  As this removal is not ``clean'', and about 
0.1 $M_\odot$  \citep[see][]{Lombardi_2006}
can remain bound to the RG stripped core, the characteristics of the WD can 
differ from those formed via a common envelope.  }
\item{ $60\%$ of CVs did not evolve via CE, which is the most common formation channel for field CVs.}
\item{Tidal captures did not play a significant role.}
\end{itemize}

\begin{table*}
\caption{Formation channels of CVs that are present in the cluster cores of different models at the age of 10 Gyr.}
\label{tab-channels}
\begin{tabular}{@{}l c c c c c c c c c c c c c c c c c c}
\hline
channel &  1a & 1b & 1c & 2a & 2b & 2c & 3a & 3b & 3c & 4a & 5a &  6a 
&\parbox[c]{8.mm}{Total}
&\parbox[c]{8.mm}{Detec}\\
\hline
standard   &  0.271 & 0.077  & 0.013     & 0.135 & 0.129  & 0.052    & 0.090 & 0.006  & 0.039    & 0.013 &    0.026 & 0.148  &  209&  47 \\
metal-rich &  0.204 & 0.056  & 0.031     & 0.148 & 0.143  & 0.046    & 0.051 & 0.015  & 0.015    & 0.020 &    0.031 & 0.230  &  265&  16\\
med-dens   &  0.327 & 0.253  & 0.167     & 0.111 & 0.012  & 0.056    & 0.031 & 0.019  & 0.006    & 0.006 &    0.000 & 0.006  &  193&  35\\
low-dens   &  0.404 & 0.066  & 0.456     & 0.037 & 0.000  & 0.007    & 0.015 & 0.000  & 0.015    & 0.000 &    0.000 & 0.000  &  156&  26 \\
fast MB    &  0.190 & 0.103  & 0.017     & 0.086 & 0.190  & 0.172    & 0.034 & 0.017  & 0.000    & 0.017 &    0.052 & 0.103  &   79&  15 \\
CE coll   &   0.212 & 0.106  & 0.006     & 0.159 & 0.194  & 0.041    & 0.041 & 0.029  & 0.029    & 0.006 &    0.000 & 0.176  &  230&  47\\ 
BF05      &   0.206 & 0.119  & 0.024     & 0.135 & 0.135  & 0.056    & 0.040 & 0.032  & 0.024    & 0.024 &    0.008 & 0.175  &  162&  36\\
47 Tuc    &   0.135 & 0.094  & 0.000     & 0.250 & 0.146  & 0.073    & 0.042 & 0.042  & 0.021    & 0.031 &    0.000 & 0.156  &  275&  37\\ 
47 Tuc+BF05 &   0.143 & 0.057  & 0.014     & 0.171 & 0.143  & 0.029    & 0.014 & 0.043  & 0.029    & 0.000 &    0.043 & 0.300  &  190&  35\\ 
47 Tuc+SCBF05 & 0.071 & 0.114  & 0.000     & 0.100 & 0.114  & 0.057    & 0.014 & 0.014  & 0.014    & 0.014 &    0.029 & 0.443  &  237&  27 \\
47 Tuc+$\alpha_{\rm CE}\lambda$  & 0.170 & 0.057  & 0.011     & 0.182 & 0.125  & 0.045    & 0.034 & 0.011  & 0.011    & 0.023 &    0.023 & 0.307  &  253&  40 \\
\hline 
non-dyn   &    &   &      &  &   &     &  &   &     &  &   &   & 124& 16\\
field     &    &   &      &  &   &     &  &   &     &  &   &   & 117& 3\\
\hline
\end{tabular}

\medskip
Notations for channels -- see text in \S~\ref{form-chan} and also Fig.~\ref{cv-scen-blue}.
``Total'' is the number of CVs and ``Detec'' is the number of detectable CVs,
both numbers are scaled per 50 000 $M_\odot$ stellar population mass in the core.
\end{table*}

\subsubsection{Formation channels in different clusters}

In Table~\ref{tab-channels} we give details on the formation channels for 
different cluster models at the same age of 10 Gyr. 
We note that these numbers fluctuate with time and are not defined precisely
(see more below in \S\ref{cv-ages}), however some trends can be identified.
We also show the numbers of CVs that are formed in a metal-poor
environment  (``non-dyn'') and in the field. The definition of ``detectable'' CVs
is not very consistent here, as observational limits for field CVs 
are not the same as for GCs 
(and much more poorly defined), 
but we use the same limits for comparison.
It can be seen that dynamics in the ``standard'' model 
leads to an increase of the total CV production by less than a factor of two.

In the case of the ``metal-rich'' model, the turn-off mass at 10 Gyr is larger than in the ``standard'' model --
there are more massive stars in the core;  the ratio between the total numbers of CVs in two models
is roughly the ratio between their turn-off masses at this age.

As was expected, the role of purely primordial CVs (channel 1a) 
decreases in importance
when density increases (see ``standard'', ``med-dens'' and ``low-dens'' models),
although their absolute number is about the same for all three models -- once a CV
is formed outside the core, it is hard to destroy it in the core.
On the other hand, the number of systems that experience CE after entering the core (1c)
 increases as the density decreases, since it is easier for pre-CE systems
to survive in a less dense environment.
The production of almost all channels via dynamical encounters decreases with 
density, except for channel 1b, where only non-strong encounters are involved.
Overall, the total number of CVs in the core does not depend strongly on the core density,
as the dynamical destruction of 
primordial binaries that would produce CVs, 
and the dynamical production
of CVs, compensate each other.

The ``fast MB'' model shows the greatest difference 
with the ``standard'' model 
in the total number of CVs that are present in the cluster core:
the ``standard'' model has about 3 times more CVs, both total and detectable, 
although the number of CVs that are ever formed in the core is slightly smaller. 
The ``fast MB'' model employs the prescription of MB with faster angular momentum loss than
in the case of the standard model, and therefore the duration of the CV stage is shorter.
 
The ``CE coll'' model does not show significant differences with our ``standard'' model. 

 The ``BF05'' model shows that CV formation is reduced mainly for primordial CVs and for CVs produced via
binary encounters. The number of CVs  produced via physical collisions between single stars
is about the same.

The results for ``47 Tuc'' are due to a mixture of several conditions: the  
higher core number density
favors dynamical formation, and the higher metallicity gives a wider
mass range over which MB operates on the donor star.
Variation of initial conditions, such as a smaller binary fraction, does not lead to significant differences
except that the relative role of binaries becomes smaller than the role of physical collisions.
Some decrease in the number of detectable CVs occurs  when we start with a smaller
initial core.

\begin{table*}
\caption{Formation channels of CVs that are present in the ``standard'' cluster core at different ages.}
\label{tab-channels-age}
\begin{tabular}{@{}l c c c c c c c c c c c c c c c c c c}
\hline
channel &  1a & 1b & 1c & 2a & 2b & 2c & 3a & 3b & 3c & 4a & 5a &  6a 
&\parbox[c]{8.mm}{Total}
&\parbox[c]{8.mm}{Det}\\
\hline
  1 Gyr & 0.250 & 0.000  & 0.000     & 0.000 & 0.250  & 0.250    & 0.000 & 0.000  & 0.000    & 0.000 &    0.000 & 0.250  &   15&   7 \\ 
  2 Gyr & 0.111 & 0.074  & 0.037     & 0.259 & 0.111  & 0.185    & 0.000 & 0.000  & 0.037    & 0.000 &    0.000 & 0.185  &   84&  37 \\ 
  3 Gyr & 0.185 & 0.046  & 0.015     & 0.154 & 0.169  & 0.123    & 0.015 & 0.000  & 0.015    & 0.031 &    0.015 & 0.231  &  170&  62 \\ 
  4 Gyr & 0.200 & 0.078  & 0.022     & 0.167 & 0.133  & 0.133    & 0.022 & 0.000  & 0.011    & 0.022 &    0.011 & 0.200  &  203&  76 \\ 
  5 Gyr & 0.190 & 0.076  & 0.029     & 0.162 & 0.086  & 0.095    & 0.057 & 0.010  & 0.019    & 0.019 &    0.019 & 0.229  &  210&  62 \\ 
  6 Gyr & 0.214 & 0.077  & 0.026     & 0.162 & 0.103  & 0.060    & 0.051 & 0.009  & 0.026    & 0.026 &    0.017 & 0.231  &  211&  56 \\ 
  7 Gyr & 0.268 & 0.049  & 0.016     & 0.163 & 0.114  & 0.041    & 0.049 & 0.008  & 0.024    & 0.024 &    0.016 & 0.220  &  204&  43 \\ 
  8 Gyr & 0.284 & 0.061  & 0.014     & 0.128 & 0.122  & 0.068    & 0.054 & 0.007  & 0.041    & 0.020 &    0.014 & 0.182  &  228&  44 \\ 
  9 Gyr & 0.268 & 0.067  & 0.013     & 0.128 & 0.134  & 0.060    & 0.074 & 0.007  & 0.047    & 0.020 &    0.013 & 0.161  &  214&  40 \\ 
 10 Gyr & 0.271 & 0.077  & 0.013     & 0.135 & 0.129  & 0.052    & 0.090 & 0.006  & 0.039    & 0.013 &    0.026 & 0.148  &  209&  47 \\ 
 11 Gyr & 0.253 & 0.084  & 0.006     & 0.157 & 0.139  & 0.054    & 0.078 & 0.006  & 0.042    & 0.012 &    0.030 & 0.139  &  212&  51 \\ 
 12 Gyr & 0.209 & 0.088  & 0.005     & 0.165 & 0.181  & 0.049    & 0.066 & 0.016  & 0.033    & 0.005 &    0.027 & 0.154  &  221&  41 \\ 
 13 Gyr & 0.203 & 0.091  & 0.005     & 0.188 & 0.162  & 0.056    & 0.066 & 0.030  & 0.030    & 0.010 &    0.025 & 0.132  &  228&  34 \\ 
 14 Gyr & 0.199 & 0.131  & 0.000     & 0.184 & 0.155  & 0.039    & 0.068 & 0.044  & 0.029    & 0.010 &    0.019 & 0.121  &  229&  43 \\ 
 \hline
\end{tabular}
\medskip

Notations are as in Table~\ref{tab-channels}.
\end{table*}

\subsubsection{Formation channels at different cluster ages}
\label{cv-ages}

\begin{figure}
  \includegraphics[height=.35\textheight]{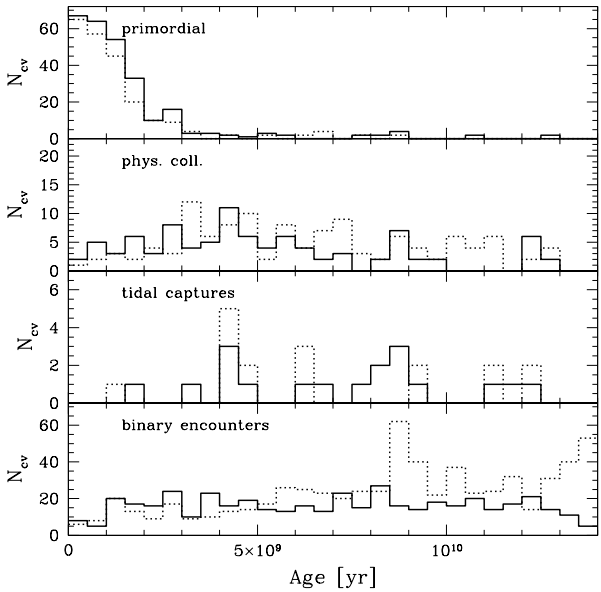}
  \caption{Formation of CVs via different channels. 
    The solid line shows the case of ``standard'' model, dotted line shows ``metal-rich''.}
\label{cv-form4w}
\end{figure}

\begin{figure}
  \includegraphics[height=.35\textheight]{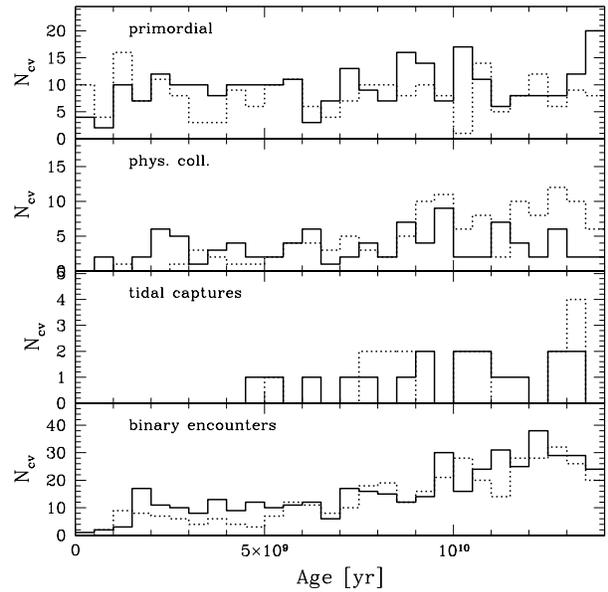}
  \caption{Appearance of CVs (time that MT starts) formed via different channels. Notations as in Fig.~\ref{cv-form4w}
}
\label{cv-appear4w}
\end{figure}

\begin{figure}
  \includegraphics[height=.35\textheight]{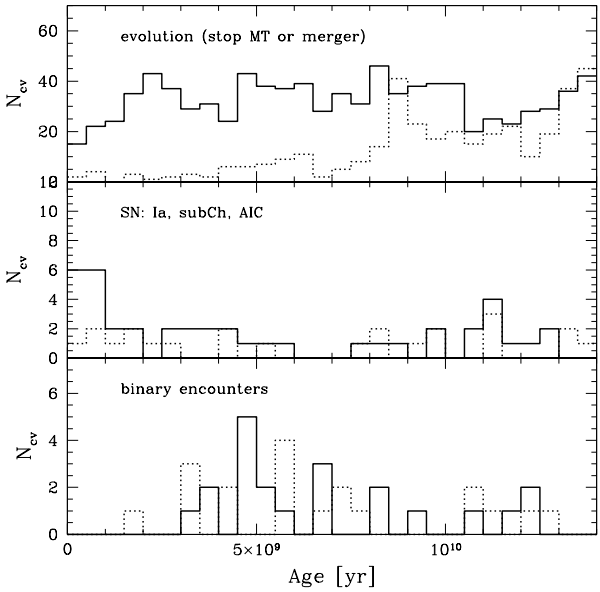}
  \caption{Destruction of CVs. Notations as in Fig.~\ref{cv-form4w}.
}
\label{cv-destr4w}
\end{figure}

In Fig.~\ref{cv-form4w}  we show the formation rate of CVs 
via different channels throughout the life of a cluster 
(indicating the time when a dynamical event occurred, or, for primordial binaries, the time when a CE happened).   
In Fig.~\ref{cv-appear4w} we show the rate of appearance of CVs (start  
of mass transfer) as a function of time.

The primordial channel produces most of its CVs at the beginning of the cluster evolution,
though the appearance of primordial CVs is  distributed flatly in time.
This contrasts with binary encounter channels, where the formation occurs
rather flatly in time, but the rate of CV appearance grows after  7 Gyr. 
A similar delay in the appearance can be seen for CVs formed via tidal captures.

In Fig.~\ref{cv-destr4w} we show the number of CVs that stop MT for different reasons:
(1) end of MT due to the star's contraction (usually occurs with CVs where a donor is a RG) or evolutionary merger of the binary; 
(2) explosion of the WD as Ia SN or sub-Chandrasekhar SN explosion, or accretion induced collapse (AIC); 
(3) end of MT due to a strong dynamical encounter. 
Most CVs stop MT due to an evolutionary reason,
while the number of SN explosions is also relatively high and is comparable
to the number of CVs destroyed by dynamical encounters 
\citep[for more detail on how SN Ia are calculated, see \S 4.4 and][]{Bel05_sn}.

In Table~\ref{tab-channels-age} we show a detailed representation of CV formation
by different channels in the ``standard'' model at different cluster ages. 
We note a peak in the number of detectable CVs at the age of 4 Gyr, and that the total number of CVs
 increases steadily until the age of 8 Gyrs and then stays constant.
The weight of different channels in the relative numbers of appearing CVs does not
change dramatically during the cluster evolution.

\subsection{Population characteristics of CVs}

\subsubsection{Periods and masses}

\begin{figure}
  \includegraphics[height=.35\textheight]{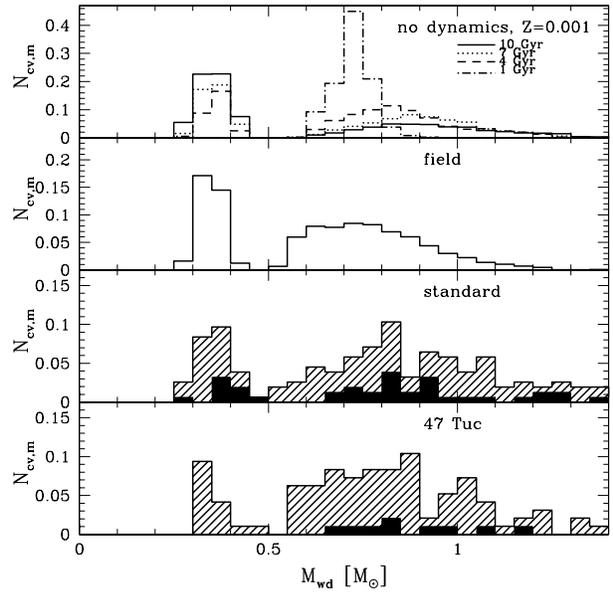}
  \caption{The mass-distribution of hydrogen accreting WDs. 
The hatched  area corresponds to dynamically formed binaries; the solid filled area to systems 
formed directly from primordial binaries. The top panel shows the case with no dynamics (different ages),
the second panel from the top shows the compiled field case, the third 
panel shows the core of the cluster in the ``standard'' model and the bottom panel shows
 the cluster core in the model ``47 Tuc''. }
\label{cv_mass}
\end{figure}

\begin{figure}
  \includegraphics[height=.35\textheight]{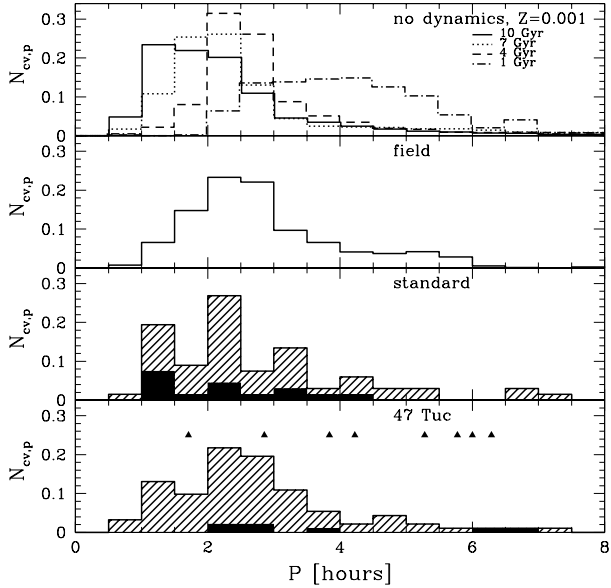}
  \caption{The period-distribution of hydrogen accreting WDs. 
The hatched area corresponds to dynamically formed binaries; the solid filled area to systems 
formed directly from primordial binaries. The top panel shows the case with no dynamics (different ages),
the second panel from the top shows the compiled field case, the third 
panel shows the core of the cluster in the ``standard'' model and the bottom panel shows
 the cluster core in the model ``47 Tuc''. Solid triangles indicates the periods of
CVs that are identified in 47 Tuc from observations. 
}\label{cv_per}
\end{figure}

On Fig.~\ref{cv_mass} and \ref{cv_per} we show mass and period distributions for ``standard'' and ``47 Tuc'' cluster models,
as well as for the population evolved without dynamics.
A remnant of the primordial population of CVs in cluster cores follows the
distribution of primordial CVs evolved without dynamics at the same age.
However the distribution of dynamically formed CVs shows signs of younger (non-dynamical) CV populations,
and also is more populated at the high mass end.
For 47 Tuc we also show the periods of the identified CVs, 
which have a distribution consistent 
 with the period distribution of the ``detectable'' CVs in our simulations.

\subsubsection{MT rates and X-ray luminosities}

\begin{figure}
  \includegraphics[height=.35\textheight]{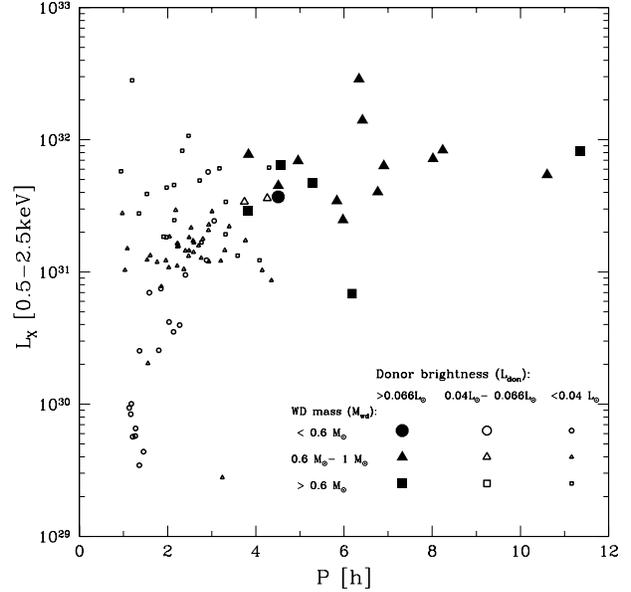}
  \caption{Simulated distribution of orbital periods versus 0.5-2.5 keV X-ray luminosity 
for ``47 Tuc'' model (at the age of 11 Gyr). 
Stars are CVs that could be detected, while open circles
are CVs that cannot be detected (generally due to the optical faintness of the donor). The size of the symbol corresponds to the WD mass
(the largest is for WDs more massive than 1 $M_\odot$, the smallest symbol is for WDs less
massive than 0.6  $M_\odot$, and the medium symbol is for WDs with masses between 0.6 and 1  $M_\odot$ ). }
\label{pl_stan}
\end{figure}

\begin{figure}
  \includegraphics[height=.35\textheight]{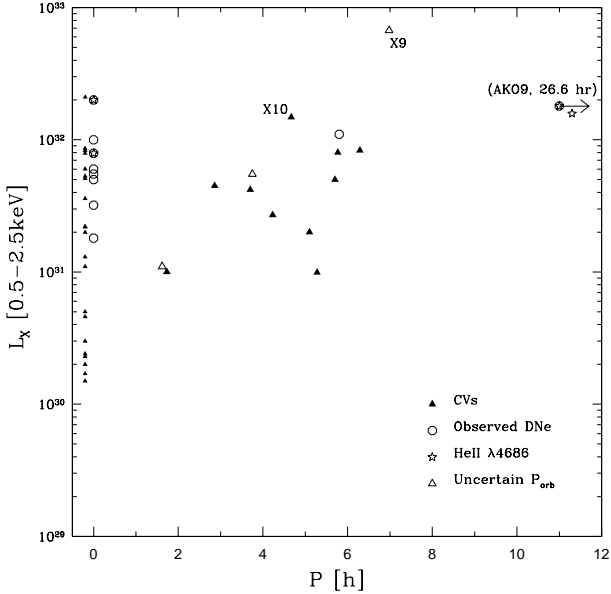}
  \caption{Observed distribution of orbital periods vs. 0.5-2.5 keV X-ray luminosity for CVs in globular clusters. 
 Known CVs in globular clusters, with known X-ray luminosities but without known orbital periods, are plotted at P=0 hours or less.  
CVs that have undergone recorded dwarf nova outbursts are indicated with circles, CVs with strong He II $\lambda4686$ emission 
(suggesting strong magnetic fields, see text) with stars (some are both).  
Three objects with uncertain periods or CV status (W34 in 47 Tuc might be a millisecond pulsar) are marked as open triangles; 
other CVs are indicated with filled triangles.  Data include 23 CVs from 47 Tuc \citep{Edmonds03a,Edmonds03b}, 
8 from NGC 6397 \citep{Edmonds99, Grindlay01b, Kaluzny03, Shara05}, 
7 from NGC 6752 \citep{Pooley02a, Bailyn96_ngc6752}, 
two from M22 \citep{Pietrukowicz05}, two from M15 \citep{Hann05}, 
one from M5 \citep{Hakala97,Neill02}, one from M4 \citep{Bassa04}, and one from M55 \citep{Kaluzny05}.}
\label{pl_obs}
\end{figure}

We compare the  distribution of orbital periods versus 0.5-2.5 keV X-ray luminosity for 
our simulation of 47 Tuc and 
observations of CVs in globular clusters 
(see Fig.~\ref{pl_stan} and Fig.~\ref{pl_obs})\footnote{Note that our ``47 Tuc'' represents a cluster similar to 47 Tuc  
but 5 times less massive}.
Observationally, few CVs have measured orbital periods.  For  
real CVs with unknown periods, only the X-ray luminosity is shown.
To obtain the X-ray luminosity of simulated CVs for comparison to observations, 
we use the accretion model from \cite{Patterson85} for 0.5-4 keV and
scale the luminosity to 0.5-2.5 keV assuming a flat energy distribution within the band:

\begin{equation}
L_{\rm x}(0.5-2.5{\rm \ keV}) = 0.066 \frac{GM_{\rm wd}\dot M}{2 R_{\rm wd}} \ ,
\end{equation}
where $\dot M$ is the mass transfer rate and $R_{\rm wd}$ is the radius of the WD.
The simulations show reasonable agreement with the observations.

However, this picture so far does not explain the rare occurrence of 
DNOs in globular clusters CVs, in comparison to the field.
Therefore,  other properties of these systems must be explored 
to identify the distinguishing characteristics.

\begin{figure}
  \includegraphics[height=.35\textheight]{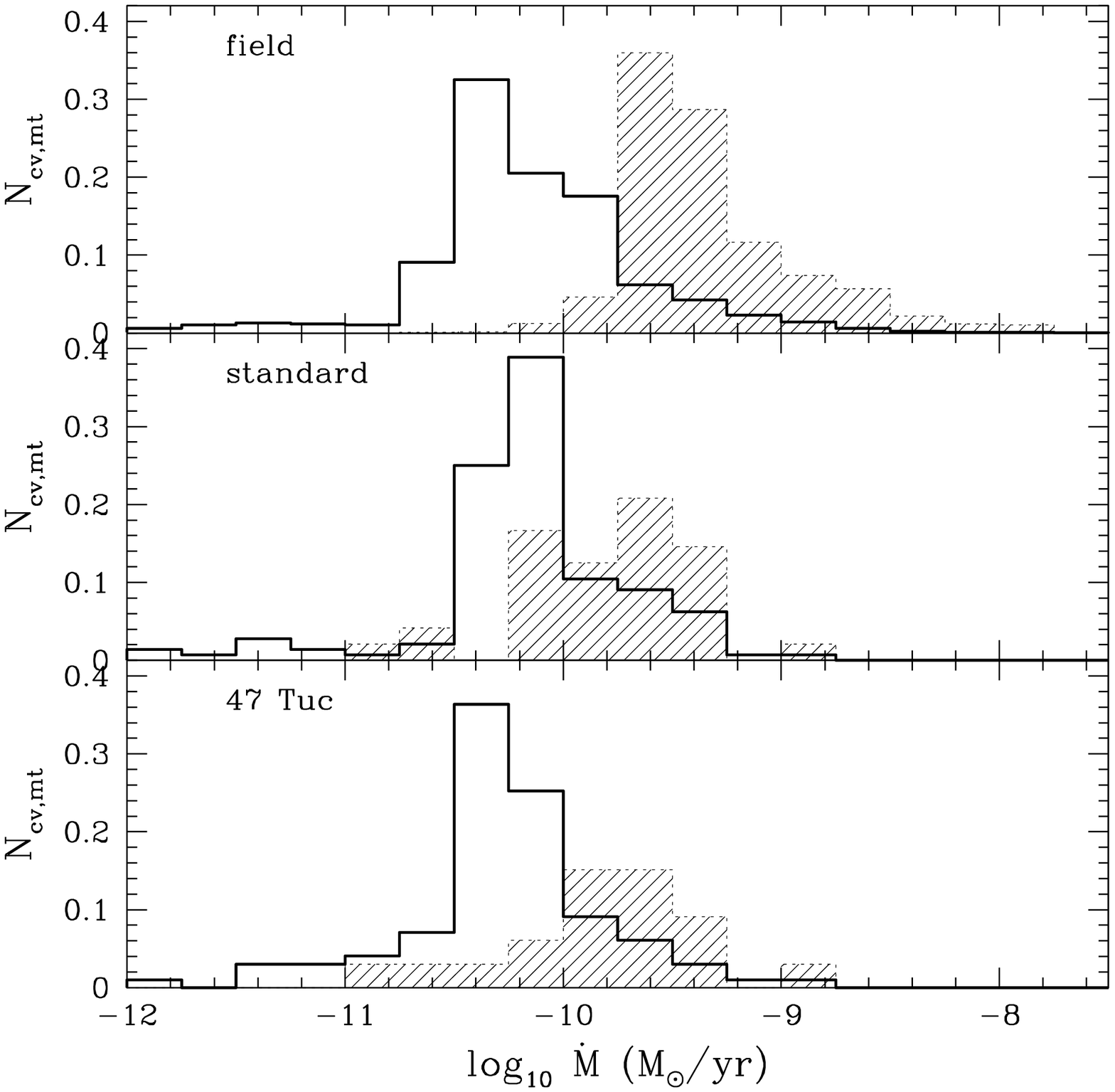}
  \caption{MT rates in CVs. Hatched areas indicate detectable CVs. For our no-dynamics model the detectable CV histogram 
is increased by a factor of 10, while for ``47  Tuc'' it is increased by a factor of 3. ``47  Tuc'' is shown at the age of 11 Gyr.}
\label{mt_cv}
\end{figure}

In Fig.~\ref{mt_cv} we show MT rates in CVs in ``standard'' model, in the 
``47 Tuc'' model and for the field. Although MT rates in our GC simulations
do not exceed $10^{-9} {\rm\,M_\odot/yr}$, and field CVs can have higher MT rates,
this result may be due to small number statistics. 
We therefore do not find significant differences between MT rates in GC CVs
and field CVs.
For our GC CVs, all MT rates are such that the accretion
disc is partially ionized and unstable,
in accordance with the disc instability model.
We adopt the viscosity parameters $\alpha_{\rm hot}=0.1$ and 
 $\alpha_{\rm cold}=0.01$, for hot and stable disc states respectively.  
Therefore, all our CVs should produce DNOs.

\begin{figure}
  \includegraphics[height=.35\textheight]{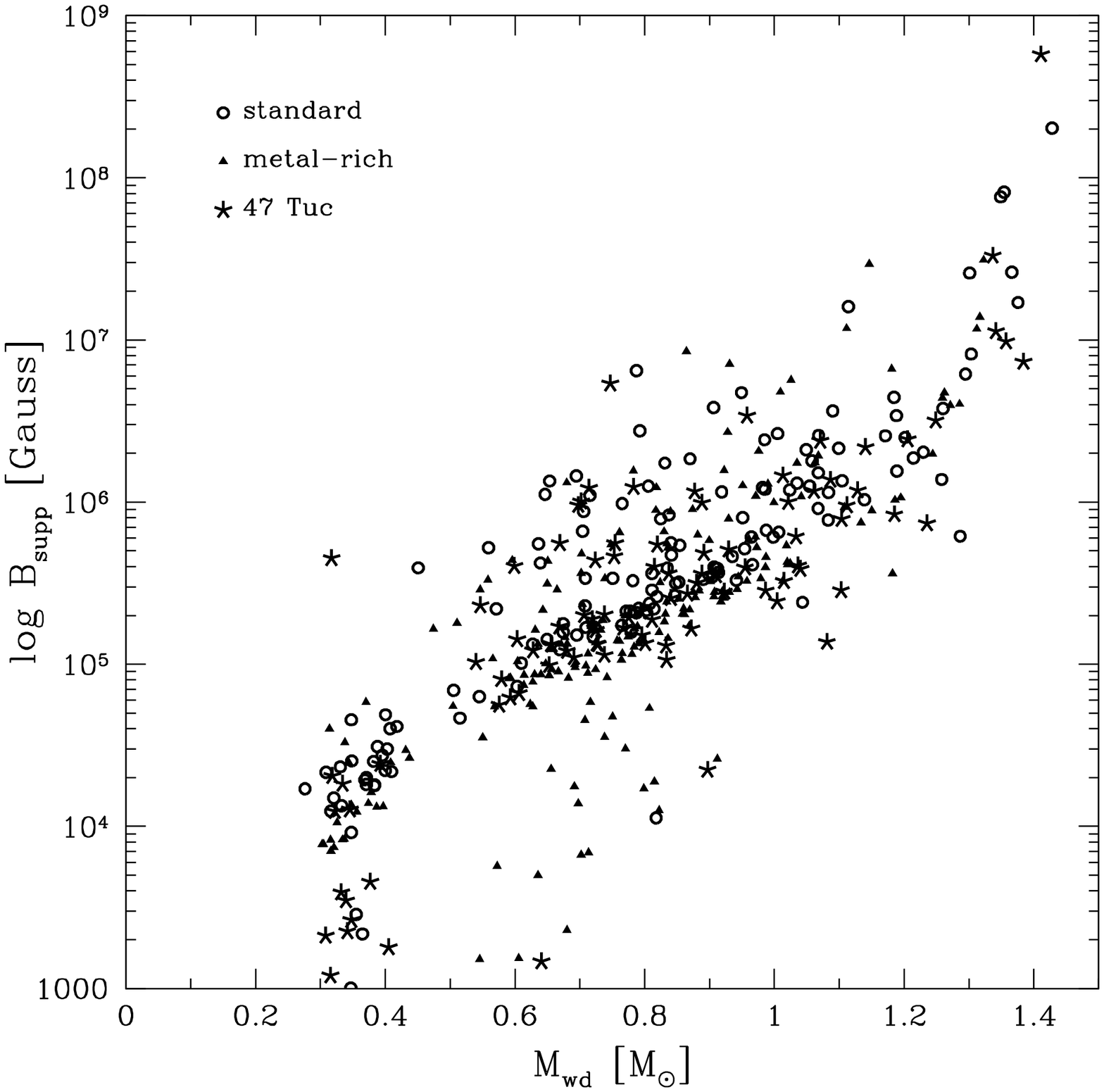}
  \caption{The minimum magnetic field $B_{\rm supp}$ required to prevent dwarf nova outbursts for CVs
in our simulations. Open circles show CVs in the ``standard'' model, filled triangles show CVs 
in the ``metal-rich'' model, and stars indicate CVs in the ``47 Tuc'' model.``47  Tuc'' is shown at the age of 11 Gyr,
other models are shown at the age of 10 Gyr.}
\label{cv_b}
\end{figure}

\cite{Dobrotka05} proposed that a combination of low MT rates and 
moderately strong magnetic fields
can explain the absence of DNOs in GCs.  Lower mass transfer rates would 
lead to a rarer occurrence of DNOs, and strong magnetic fields would lead to 
truncation of the inner disc keeping the disc in the cold stable state. 
As we do not find systematically lower MT rates for our GC CVs, (in fact 
our MT rates are two orders of magnitude higher than found in \cite{Dobrotka05}), we 
estimated the minimum magnetic field required to suppress DNOs 
(see Fig.~\ref{cv_b}), using the criterion from (\cite{Dobrotka05}):

\begin{equation}
B_{\rm supp} \ga 5.7\times 10^5 {\rm G}  \left ( \frac{\dot M}{10^{-10} \frac{M_\odot}{yr}} \right )^{1.16} 
\left (\frac{M_{\rm wd}}{M_\odot} \right) ^{0.83} \left (\frac{0.01R_{\odot}}{R_{\rm wd}} \right) ^{3} . 
\end{equation}
\noindent We find that $B_{\rm supp}$ is a slowly increasing function of the WD mass and, for
most of the WDs with  masses below 1 $M_\odot$, is even below $10^6$ G.
WDs with $B\la 10^6$ G are not regarded as highly magnetic. 
It can also be seen that a $10^7$ G field is enough to prevent DNOs in all
CVs with the WDs less massive than  1.1 $M_\odot$ and the field of $~10^8$ G 
is strong enough to stop DNOs for WDs of all masses.

\subsection{Formation Channels of AM~CVns}

\label{sec-amcv}

\begin{table*}
\caption{Formation channels of AM~CVns that are present in the ``standard'' cluster core at 10 Gyr.}
\label{tab-channels-amcv}
\begin{tabular}{@{}l c c c c  c c c c c  c c c c c}
\hline
channel &  1a & 1b & 1c & 2a & 2b & 2c & 3a & 3b & 3c & 4a & 5a &  6a 
&\parbox[c]{8.mm}{Total}\\
\hline
standard   & 0.385 & 0.068  & 0.000      & 0.051  & 0.090  & 0.030     & 0.013  & 0.013 & 0.064    & 0.064 &    0.009  & 0.265  &    316 \\
metal-rich & 0.379 & 0.089  & 0.000      & 0.031  & 0.071  & 0.022     & 0.018  & 0.004 & 0.009    & 0.009 &    0.000  & 0.348  &    303 \\
med-dens   &  0.732 & 0.141  & 0.056      & 0.010  & 0.010  & 0.015     & 0.005  & 0.000 & 0.000    & 0.000 &    0.000  & 0.076  &    236 \\
low-dens   &  0.680 & 0.085  & 0.178      & 0.008  & 0.004  & 0.036     & 0.000  & 0.000 & 0.000    & 0.000 &    0.000  & 0.178  &    283  \\
fast MB    & 0.450 & 0.077  & 0.000      & 0.036  & 0.068  & 0.045     & 0.009  & 0.005 & 0.027    & 0.027 &    0.000  & 0.266  &    303\\
CE coll   &  0.474 & 0.123  & 0.000      & 0.035  & 0.082  & 0.023     & 0.006  & 0.012 & 0.018    & 0.018 &    0.012  & 0.211  &    231 \\ 
BF05      &   0.271 & 0.090  & 0.005      & 0.053  & 0.080  & 0.016     & 0.011  & 0.032 & 0.011    & 0.011 &    0.000  & 0.404  &    242 \\
47 Tuc    &   0.246 & 0.070  & 0.000      & 0.026  & 0.070  & 0.035     & 0.035  & 0.018 & 0.009    & 0.009 &    0.000  & 0.491  &    327\\ 
47 Tuc+BF05 &   0.174 & 0.110  & 0.000      & 0.018  & 0.101  & 0.037     & 0.000  & 0.009 & 0.009    & 0.009 &    0.009  & 0.514  &    296 \\
47 Tuc+SCBF05&  0.217 & 0.101  & 0.000      & 0.029  & 0.043  & 0.043     & 0.000  & 0.000 & 0.014    & 0.014 &    0.000  & 0.551  &    233\\
47 Tuc+$\alpha_{\rm CE}\lambda$ & 0.252 & 0.078  & 0.000      & 0.000  & 0.117  & 0.029     & 0.019  & 0.010 & 0.058    & 0.058 &    0.000  & 0.427  &    296\\
\hline
non-dyn  & & & & & & & & & & & & & 169\\
field  & & & & & & & & & & & & & 110\\
\hline
\end{tabular}
\medskip

Notations for channels is the same as for CVs (see text in \S~\ref{form-chan} and also Fig.~\ref{cv-scen-blue}).
``Total'' is the number of AM~CVns, the number is scaled per 50 000 $M_\odot$ stellar population mass in the core.
\end{table*}

\begin{figure}
  \includegraphics[height=.35\textheight]{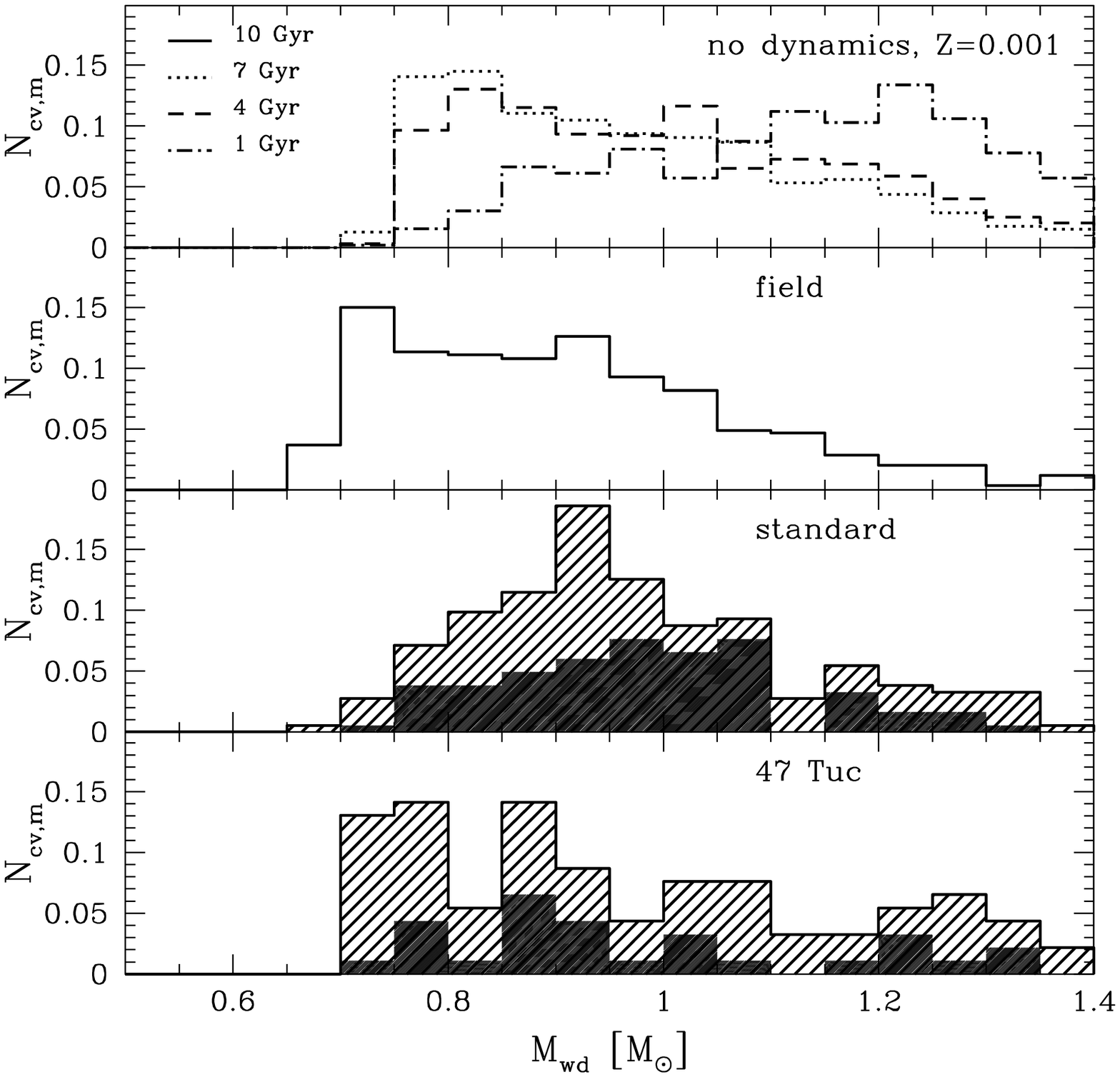}
   \caption{The mass-distribution of helium accreting WDs. 
The hatched  area corresponds to dynamically formed binaries; the solid filled area to systems 
formed directly from primordial binaries. The top panel shows the case with no dynamics (different ages),
the second panel from the top shows the compiled field case, the third 
panel shows the core of the cluster in the ``standard'' model and the bottom panel shows
 the cluster core in the model ``47 Tuc''. }
\label{amcv_mass}
\end{figure}

In Table~3 we show the main formation channels for AM~CVns that occur in different clusters,
classifying channels in an analogous manner to Sec.~4.1.1.
The main differences with the formation of CVs are: (i) post-CE channels (including primordial) are more important;
(ii) the role of physical collisions for AM~CVns  is increased;
(iii) binary encounters played a much less significant role.

In most physical collisions the participants are a RG of $0.9-2.2\,M_\odot$
(the core mass is $0.2-0.3\,M_\odot$) and a WD $\ga 0.6\,M_\odot$.
In $\la 20\%$ of physical collisions the participants 
are a He core burning giant of $1.7-2.2\,M_\odot$ (the collision in this case leads to the formation
of a He star of $0.4-0.55\,M_\odot$)  and a heavier WD, generally $\ga 0.9-1.1\,M_\odot$.

Overall, this channel provides AM~CVns with accretors of masses 
from $0.8\,M_\odot$ to $1.4\,M_\odot$ at a cluster age of 10 Gyr.
The field population of AM~CVns would have accretors with masses typically between $0.65\,M_\odot$ and $1.0\,M_\odot$ 
\citep[Fig.~\ref{amcv_mass}, see also ][]{Nelemans_2001_amcvns,Bel05_lisa}.
The peak in the accretor mass distribution is shifted from 
$\sim 0.7\,M_\odot$ in the field population to $0.9\,M_\odot$ in the cluster population.

\subsection{Explosive events}

\label{sec-expl}
 
\begin{table}
\caption{Rate of explosive events.}
\label{tab-channels-expl}
\begin{tabular}{@{}l c c c c c c}
\hline
event &   SN Ia  & supraCh & subCh & AIC & NS$_{\rm DD}$ & NS$_{\rm AIC}$\\
\hline
standard   &   2.0 &   9.3 &   7.3 &   3.21 & 118 &  79.2\\
metal-rich &    3.48 &   8.2 &  11.7 &  2.73 & 117 &  91.4\\
med-dens   &     0.48 &   2.4 &   8.0 &  1.45 &  89 &  64.7 \\
low-dens   &     0.00 &   2.9 &   9.8 &  0.72 &  80 &  61.3 \\
fast MB    &      3.48 &   8.2 &  11.7 &  2.73 & 117 &  91.4\\
CE coll   &     0.74 &   7.7 &   8.9 &  2.97 & 117 &  85.1\\ 
BF05      &     1.56 &   3.6 &   4.2 &  2.45 &  70 &  45.5\\
47 Tuc    &    1.22 &   4.9 &   9.3 &  2.44 &  88 &  70.3\\ 
47 Tuc+BF05 &     1.54 &   2.2 &   6.4 &  0.66 &  65 &  42.1\\
47 Tuc+SCBF05&    0.66 &   2.4 &   6.2 &  0.88 &  50 &  36.3\\
47 Tuc+$\alpha_{\rm CE}\lambda$ &    2.20 &   3.2 &   8.8 &  1.71 &  87 &  59.6 \\
\hline
non-dyn  &    0.00 &   0.7 &   3.2 &  0.00 &  70 &  69.4  \\
field   &   0.50 &  7.7 &  15.1 &   1.58 & 122.17  &   22.8\\
\hline
\end{tabular}
\medskip

``SN Ia'' is the number of Type Ia SN   (single-degenerate channel only), 
``supraCh'' is the number of double WD mergers where the total mass is more than
$1.4\,M_\odot$, ``subCh'' is the number of sub-Chandrasekhar nuclear runaways,
``NS'' is the number of NSs that can be potentially formed via double-degenerate (NS$_{\rm DD}$) 
or AIC (NS$_{\rm AIC}$) channels 
until the age of 10 Gyrs. For cluster models, rates and numbers are given per Gyr per 200,000 $M_\odot$ total cluster mass
and are averaged for the ages of 8-12 Gyr. 
For non-dynamical models, rates and numbers are given for the age of 10 Gyr and for the field model
rates and numbers are given after 10 Gyr of continuous star formation.
\end{table}

As the mass distribution of accreting WDs is shifted towards higher masses compared 
to the field population, it is important to check what effect this has on the rates of:
(i) Type Ia supernova (SN Ia, here we mean the single degenerate channel only);
(ii) double WD mergers (those for which the total mass $\ge M_{\rm Ch}\simeq 1.4\,M_\odot$);
(iii) sub-Chandrasekhar supernovae;
and (iv) accretion induced collapse (AIC).

The type of event that occurs depends on the
mass and composition of the white dwarf and the rate of mass transfer.
If this is a carbon-oxygen (CO) WD and experiences stable accretion, it will accumulate mass until
it reaches the Chandrasekhar limit and then explodes as a type Ia supernova.
In the case of specific MT rates the accretion leads to the accumulation of He in the shell
\citep{Kato99_ia}. 
If sufficient mass is accumulated, it will lead to the ignition of the CO or ONeMg core
and disruption of the WD as a sub-Chandrasekhar-mass Type Ia supernova
\citep[see][]{Taam80_subch,Livne91_subch, Woosley94_subch,Garcia99_subch, Ivanova_04d}.
In the case of accretion on to ONeMg WDs, upon reaching the Chandrasekhar limit
the WD will undergo  accretion-induced collapse and form a NS.

If the donor is another WD and the mass transfer is not stable, 
the mass of the merger product can exceed the Chandrasekhar limit  -- 
these so-called supra-Chandrasekhar mergers could lead
either to a Type~Ia supernova (double-degenerate channel), or to a merger induced
collapse of the remnant to form a NS and perhaps a millisecond radio pulsar \citep{Chen93}.
It was argued as well that in the latter case and, if one of the WD is magnetic,
such mergers  will lead to magnetar formation.  Such objects may be 
responsible for the production of the giant flares emitted by soft 
$\gamma$-repeaters, which can be identified with early type galaxies. These
flares may contribute to a fraction of the observed short duration burst 
population at higher redshift \citep{Levan06_SGR}.

If NSs are born with natal kicks, most of them will be ejected 
from the shallow cluster potential, leaving few NSs to explain
the observed number of millisecond pulsars \citep{Pfahl02}.
In the case of accretion or merger induced collapse, the NSs are likely to be 
formed without a significant kick \citep{Podsi04_aic},  
and this can relieve the NS ``retention problem''. 
If double WD mergers do not lead to collapse, they must contribute 
to the rate of Type~Ia supernovae, with potential
 cosmological implications \citep[for a review see][]{Leibundgut01}.

The production of supra-Chandrasekhar mergers in our galaxy was discussed
in \cite{Hurley02} and was estimated to be 2.6 per year in the Galactic disc
(this is 8.6 per cluster per Gyr in our units).
Several free parameters can have strong effects on this result, such as the 
common envelope prescription, the initial mass function,
or the adopted star formation history. 
There are also differences between our models
and those of \cite{Hurley02}: 
(i) their cut-off mass for WD binaries is at the initial mass
of $0.8\, M_\odot$; 
(ii) they choose $\alpha_{\rm CE}=3$; 
(iii) they adopted continuous star formation through 15 Gyr (c.f. ours 10 Gyr); and 
(iv) our model for accretion on to WDs 
is more up-to-date \citep[for details, see][]{Bel05b}.
Overall, we find that our formation rates for the field are not significantly
different from those of \citet{Hurley02} (see Table~\ref{tab-channels-expl}). 
We also find that our rates are smaller if the star formation is taken not as flat, but with
one (or several) star formation bursts that ended several Gyrs ago.

The enhanced production rate of double WD mergers in dense stellar clusters
was first discussed in detail by \citet{Shara02}, who applied this to 
open clusters. They found that the supra-Chandrasekhar WD merger rate 
can be increased by an order of magnitude (although their statistics were based on only a few events).
We did not find such an increase compared to the field population, 
where star formation
is continuing, though we found some increase compared to the case without dynamical interactions  
(see Table~\ref{tab-channels-expl}). 
However, we note that our total 
number of supra-Chandrasekhar WD mergers is large.
In fact, if indeed all those mergers lead to formation of NSs,
and those NSs are retained by the cluster, then this channel 
provides about 6\% of all NSs ever created. 
The NSs thus created become comparable in numbers to
the NSs that were born with natal kicks and retained.
The production of NSs via this channel can be reduced 
by reducing the efficiency of the common envelope. 
In this case, 
more binaries will merge during the CE phase 
and less supra-Chandrasekhar mergers will occur.
We found however that even the reduction 
of $\alpha_{\rm CE}\lambda$ to 0.1 led only to a moderate decrease of the ``current'' (at about 10 Gyr) 
production rate of supra-Chandrasekhar mergers, while their total production is only a bit smaller
(see different models for 47 Tuc in the Table~\ref{tab-channels-expl})).
In addition, the production of NSs via AIC is comparable to the
production of NSs via merger induced collapse, and therefore also appears to be
 a significant source of NSs in GCs.  
The question of how many NSs can be created via different channels
in GCs is very important, and will be addressed in more detail in Paper II.

We estimate the contribution 
 of SN Ia produced in GCs to total galactic SN rates.
Assuming that  $\sim 3\times 10^{7}\,M_\odot$ 
 is contained in galactic GCs ($\sim 150$ galactic GCs),
we find that the single-degenerate channel from GCs can provide 
at best 1 SN per $\sim 10^6$ yr per galaxy,
and the contribution of GCs is only several times higher if 
the double-degenerate channel (supra-Chandrasekhar) also results in SN. 
In spiral galaxies the rate of SN Ia is  0.04-0.1 per century per galaxy \citep{Mannucci05_Iarate}, 
and therefore the contribution of GCs is not important.
However, GCs can play a larger role in elliptical galaxies, where no star formation 
is going on and the rate of SN Ia provided by the field population is smaller.
In addition, in ellipticals the specific frequency of GCs per galaxy luminosity unit is significantly higher than
in spirals \citep[up to 8.6 compared to 0.5 in Milky Way, see e.g.][]{Kim04_xlf}.
Also, it has been shown that the observational SN Ia rate consists 
of two components -- a prompt piece that is proportional to the star formation rate, 
and an extended piece that is proportional to the total stellar mass \citep{Scannapiec05_Ia}. 
This is consistent with the behavior of the formation rates of both single degenerate 
and double degenerate channels in GCs, which 
peak during the first Gyr of the cluster evolution and 
have a flat distribution at later times.
We therefore propose that GCs can increase the theoretically predicted rates of SN Ia in 
elliptical galaxies.

\subsection{Coalescing compact binaries as LISA sources}

\begin{figure}
  \includegraphics[height=.35\textheight]{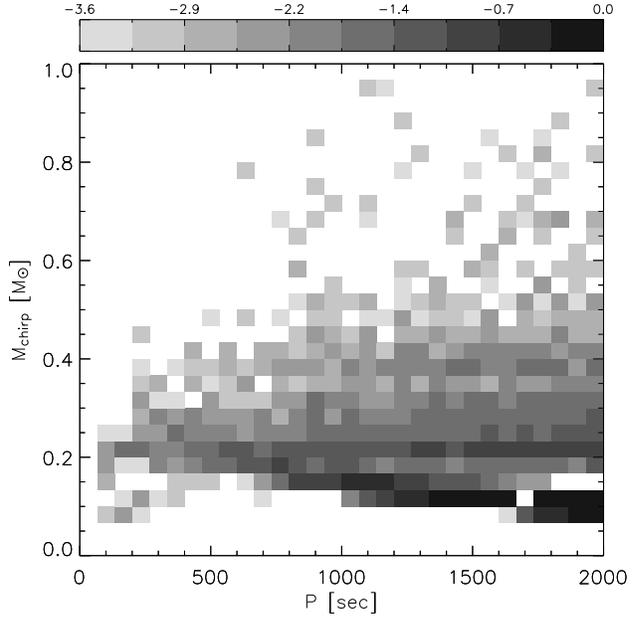}
  \caption{Distribution density (averaged over time) of LISA binaries, 
in the space of binary periods and chirp masses, from our ``standard'' model 
(integrated over cluster ages from  9 to 13 Gyr).}
\label{lisa-bin}
\end{figure}

AM~CVns discussed in \S\ref{sec-amcv}, as well as most double white dwarf mergers 
discussed in \S\ref{sec-expl} (with the exclusion of a small fraction of mergers
that are results of physical collision during hard binary encounters)  
are coalescing binaries driven by gravitational radiation. 
Prior to their mergers, or before and during the MT, they can be detectable as gravitational wave sources by LISA. 
Their detectability is significantly enhanced when their orbital periods become smaller than about $2000\,$s
\citep{Ben01,Nelemans05_lisa}, so their signals can be distinguished from the background noise
produced by Galactic binaries. As the positional accuracy of
LISA will be much greater for binaries with such short periods, these sources
can be associated with specific globular clusters in our Galaxy.

From our simulations we find that at any given moment, a typical cluster of 200,000 $M_\odot$
will contain 10 LISA sources on average,
and at least 3 LISA binaries at any given moment; during 1 Gyr a cluster forms 180 LISA systems. 
A massive cluster like 47 Tuc, with mass $\sim 10^6\,M_\odot$, 
will have at least 10 LISA sources at any given moment, and 40 LISA binaries on average.
At an age of 10 Gyrs such a cluster can produce as many as 3-15 NS-WD LISA binaries per Gyr
(a typical cluster produces 1-3 NS-WD coalescing binaries per Gyr).

With the total mass in GCs of about $ 3\times 10^{7}\,M_\odot$, as many as 1500 LISA binaries
can be present in all Galactic GCs (this number will decrease
if the CE efficiency is smaller).
The most optimistic upper limit for the galactic formation 
rate of NS-WD binaries in GCs is several hundred per Gyr. 
The lifetime of a NS-WD binary in the LISA-band during the MT is $\sim 10^8$ 
yr. The time prior to the onset of MT depends on the binary eccentricity, but is usually much shorter \citep{Ivanova05_ucxb}.
With our predicted 
formation rates, as many as 10-50 NS-WD LISA binaries 
can be detected in GCs. This number is probably too optimistic,
as fewer than 10 ultra-compact X-ray binaries (mass-transferring NS-WD binaries with 
orbital periods less than an hour) have been identified in GCs,
implying that the formation rate of LISA-sources should be smaller. 
(One explanation could be that 
 many globular clusters are less dense than our ``standard'' model and 
NS-WD formation is thus less frequent.) We shall address the issue of the 
formation rates of binaries with NSs in more detail in Paper II. 
Here we will only note that the LISA binaries will spend
spend most of their time in the LISA band among their MT tracks, 
with chirp masses $\la 0.25 M_\odot$ (see Fig.~\ref{lisa-bin}).

\section{Discussion}

With our simulations we predict that the formation rates of CVs and AM~CVns
in globular clusters are not very different from those in the field population.
The numbers of CVs and AM CVns per mass unit are comparable to 
numbers in the field  if the whole 
cluster population is considered, 
and they are only 2-3 times larger in the core than in the field.
Dynamical formation is responsible only for 60\%-70\% of CVs
in the core. This fraction decreases as the density 
decreases, and 
the role of primordial CVs becomes more important.
We rule out tidal captures as an effective mechanism 
for CV formation in GCs unless the rate
of TCs is significantly underestimated. Instead we propose that the population of GC CVs reflects
a combination of primordial CVs, CVs in post-exchange binaries, and 
products of physical collisions of MS stars with RGs. There
are also primordial CVs which are located in the halo and have 
never entered the core.
The GC core density variation indeed does not play a 
very large role, in contrast to the
case of NS binaries, where almost all systems are formed dynamically \citep{Ivanova04a} and whose numbers have
a strong dependence on the cluster collision rate \citep{Pooley03}.
We expect to have one detectable CV per $1000\,M_\odot$ in the core of a typical cluster and
about one detectable CV per $1000-2000\,M_\odot$ in a 47 Tuc type cluster.  
Thus we predict 35-40 CVs in the core of 47 Tuc, 
in quite reasonable agreement with observations, where 22 CVs in 47 Tuc 
have been 
identified \citep{Edmonds03a}.  Even better agreement 
between our simulations and the observed number of CVs can be obtained if we assume that
the initial core mass in 47 Tuc is smaller than 5\% of the cluster.

Although the formation rates do not differ strongly, we
found significant differences in the populations, and note that 
these differences may have observational consequences.
Indeed, cataclysmic variables in globular clusters have an unusual array of 
characteristics, which make them difficult to classify as members of the 
standard classes of CVs recognized in the galaxy.  Their X-ray luminosities
seem to be rather high, compared with CVs in the field \citep{Verbunt97}.  
They exhibit dwarf nova outbursts only rarely, compared to well-studied 
dwarf novae: only 1 dwarf nova was found in 47 Tuc by \citet{Shara96} in a 
survey which would have detected 1/3 of known dwarf novae if they were 
located in 47 Tuc, while \citet{Edmonds03a} identify 22 firm CVs in 47 Tuc.  
Finally, the X-ray to optical flux ratios of CVs in globular clusters are 
relatively high, comparable to those of dwarf novae \citep{Edmonds03b}.

One solution to this problem was the suggestion that CVs in globular clusters 
tend to be primarily magnetic in nature, compared to CVs in the field 
\citep{Grindlay95}. Magnetic CVs have no discs (AM Her or polar CVs), 
or truncated discs (DQ Her CVs or intermediate polars, IPs), because of 
the effect of the WD magnetic field.  As a result, the 
disc instability is nonexistent or suppressed.  Magnetic CVs are 
believed to produce X-rays through an accretion shock above the 
polar cap, producing high X-ray 
luminosities, while nonmagnetic CVs produce an optically thick boundary 
layer, saturating their X-ray emission \citep{Patterson85}.  
Strong He II $\lambda$ 4686 \ lines were observed in the spectra of three CVs 
in NGC 6397 \citep{Edmonds99}, indicating a strong source of FUV radiation. 
This FUV radiation could indicate either evidence for an intermediate polar 
interpretation, or a very high mass transfer rate; the second interpretation 
is favored for the FUV-bright, 27-hour period CV AKO9 in 47 Tuc 
\citep{Knigge03}.  Another argument in favor of the intermediate polar 
interpretation is the excess $N_H$ (in addition to that expected along the line of sight) observed towards many CVs in 47 Tuc 
\citep{Heinke05a}. Excess $N_H$ in CVs that are not observed at high inclinations has been considered a signature of the accretion curtains observed in the magnetic systems known as intermediate polars.

However, only two globular cluster CVs have shown clear 
evidence of magnetic fields in their X-ray lightcurves so far 
\citep[X9 and X10 in 47 Tuc][]{Grindlay01a,Heinke05a}. This may not mean 
that these systems are not magnetic, since the number of X-ray photons 
detected from globular clusters is generally small (compared with nearby, 
well-studied CVs).  In addition, it has been suggested that the accretion in 
SW Sex and VY Scl CVs is governed by the WD magnetic field, without evidence 
of pulsations \citep{Rodriguez-Gil01,Hameury02}.
 Another problem for 
the magnetic interpretation is that IPs tend to be optically brighter
than typical CVs in globular clusters, which 
have lower ratios of X-ray to 
optical flux more typical of dwarf novae than IPs \citep{Edmonds03b}.  A final 
problem is the observation of dwarf nova outbursts in two of the three CVs in 
NGC 6397 possessing strong He II emission \citep{Shara05}.  
A proposed resolution to these problems is a combination of a low mass 
transfer rate (which will reduce the optical brightness and increase the 
X-ray to optical flux ratio) with an intermediate polar nature 
\citep{Edmonds03b}. \citet{Dobrotka05} calculated the dwarf nova recurrence 
times for CV discs with various mass transfer rates and WD magnetic moments,
and found a parameter space that fulfilled the requirements of 
globular cluster CVs.  Left unanswered was why globular cluster CVs might 
tend to have stronger magnetic fields than field systems.

Our work provides a possible answer to this question.  Globular cluster 
dynamics has a strong effect on the composition of the binaries that form 
CVs, tending to place more massive WDs into binaries that will become CVs
(Fig.~11). 
Increasing the mass of WDs in CVs increases the energy that can be 
extracted at a given mass transfer rate, thus increasing the X-ray luminosity 
and X-ray to optical flux ratio of the CVs.  This effect is complementary to 
the effects of higher magnetic fields.  However, higher mass WDs also 
have a higher probability of showing strong magnetic fields 
\citep{Vennes99,Liebert03,Wick05}.  
Thus  the dynamical origin of WDs in globular cluster CVs may be responsible 
for the observational peculiarities of globular cluster CVs; their 
relatively high X-ray luminosities and X-ray to optical flux ratios, and 
their low rates of dwarf nova outbursts.  

The tendency for higher mass white dwarf accretors in GC CVs in 
comparison to the field also affects 
the production of the superhump phenomenon.  This behavior results from the precession of 
the outer disc due to the excitation of resonances 
within the disc caused by the 3:1 commensurability of  motions in the 
disc with the companion's orbital period \citep[see][]{Whitehurst91}.  
Such systems are characterized by mass ratios (of donor to accretor) of less 
than 0.25-0.33.  Systems of this type in the field are rarely observed 
at orbital periods above the period gap, but the higher white dwarf 
masses of CVs in GCs would increase their likelihood in GCs.
 
We examined also several other consequences of having a dynamically modified 
 population of close binaries including WDs. In particular, considering 
supra-Chandrasekhar mergers, we found that too many NSs may be formed
if these mergers lead to merger-induced collapse.
We suggest that either this mechanism does not lead to NS formation, 
or the CE efficiency is overestimated.
By our estimates, GCs do not contribute strongly to the SN Ia rates in spiral galaxies,
however they 
may significantly increase these rates in 
elliptical galaxies.
We have also shown that GCs can be excellent sites for LISA observations since 
many GCs will contain several LISA sources at any given moment, 
although most of those systems will have low chirp masses.

\section*{Acknowledgments}

This work was supported by NASA Grant NAG5-12044 (FAR), NSF Grant  AST-0200876 (RET), and
Chandra Theory Grant TM6-7007X (JF) at Northwestern University,
KB acknowledges supprot by KBN grant 1P03D02228.
All simulations were performed on the McKenzie cluster at
CITA which was funded by the Canada Foundation for Innovation. 

%\bibliography{Ivanova}
%\bibliographystyle{mn2e}

\end{document}